\documentclass[11pt,preprint]{aastex}
\usepackage{epstopdf}
\usepackage{epsf}
\usepackage{amsmath}
\usepackage{graphicx}
\usepackage{natbib}

\def\wisk#1{\ifmmode{#1}\else{$#1$}\fi}

\def\lsim {\wisk{_<\atop^{\sim}}}
\def\gsim {\wisk{_>\atop^{\sim}}}

\def\Msun {\wisk{{\rm M_\odot}}}
\def\arcmin {\wisk{^\prime}}
\def\arcsec {\wisk{^{\prime\prime}}}
\def\micron {\wisk{\mu{\rm m}}}

\def\um {$\mu$m}

\def\half {\wisk{\frac{1}{2}}}

\def\nwm2sr {\wisk{\rm nW/m^2/sr\ }}
\def\nw2m4sr {\wisk{\rm nW^2/m^4/sr\ }}
\def\nwm {\wisk{\rm ~nW/m^{2}/sr }}
\def\cgs {\wisk{\rm erg/cm^2/s/sr\ }}
\def\flux {\wisk{\rm erg/cm^2/s\ }}

\slugcomment{}

\shorttitle{}
\shortauthors{Cappelluti, N. et al.}

\begin{document}

\title{Cross-correlating cosmic IR and X-ray background fluctuations: evidence of significant black hole populations among the CIB sources}
%
%
%
%
%
\author{N. Cappelluti\altaffilmark{1,2}, A. Kashlinsky\altaffilmark{3,4} , R. G. Arendt\altaffilmark{3,2} , A. Comastri\altaffilmark{1} ,G. G. Fazio\altaffilmark{5}\\
A. Finoguenov\altaffilmark{6,2} , G. Hasinger\altaffilmark{7}, J. C. Mather\altaffilmark{3,8}, T. Miyaji\altaffilmark{9} \and S. H. Moseley\altaffilmark{3,8}}
\altaffiltext{1}{INAF-Osservatorio Astronomico di Bologna, Via Ranzani 1, 40127 Bologna, Italy}
\altaffiltext{2}{University of Maryland, Baltimore County, 1000 Hilltop Circle, Baltimore, MD 21250, USA}
\altaffiltext{3}{Observational Cosmology Laboratory, Code 665, Goddard Space Flight Center, Greenbelt MD 20771}
\altaffiltext{4}{SSAI}
\altaffiltext{5}{Harvard Smithsonian Center for Astrophysics, 60 Garden Street, Cambridge, MA 02138, USA}
\altaffiltext{6}{Department of Physics, University of Helsinki, Gustaf HŠllstršmin katu
2a, FI-00014 Helsinki, Finland}
\altaffiltext{7}{Institute for Astronomy, University of Hawaii, 2680 Woodlawn Drive, Honolulu, HI 96822, USA}
\altaffiltext{8}{NASA}
\altaffiltext{9}{Instituto de Astronom\'ia, Universidad Nacional Aut\'onoma de M\'exico, Km 103 
 Carret. Tijunana-Ensenada, Ensenada, 22860, BC, Mexico}

\begin{abstract}
 In order to understand the nature of the sources producing the recently uncovered CIB fluctuations, we study cross-correlations
between the fluctuations in the source-subtracted Cosmic Infrared Background (CIB) from {\it Spitzer}/IRAC data 
and the unresolved Cosmic X-ray Background (CXB) from deep {\it Chandra} observations.
Our study uses data from the EGS/AEGIS field, where both datasets cover an $\simeq 8^\prime\times 45^\prime$ region of the sky.
Our measurement is the cross-power spectrum between the IR and X-ray data. 
 The  cross-power signal between the IRAC maps at 3.6\um\ and 4.5\um\ and the Chandra [0.5-2] keV data has been detected,
 at  angular scales $\gsim 20\arcsec$,
  with an overall significance of  $ \simeq 3.8 \sigma$ and $\simeq 5.6 \sigma$, respectively. At the same time we find no evidence of significant 
  cross-correlations at the harder {\it Chandra} bands. 
The cross-correlation signal is produced by individual IR sources with 3.6\um\ and 4.5\um\ magnitudes $m_{AB}\gsim $ 25-26 and [0.5-2] keV 
X-ray fluxes $\ll 7\times 10^{-17}$ \flux.
We determine that at least $15-25\%$ of the large scale power of
the CIB fluctuations is correlated with the spatial power spectrum of the X-ray fluctuations.  
If this correlation is attributed to emission from accretion processes at both IR and X-ray wavelengths, this implies a much higher fraction 
of accreting black holes than among the known populations.
We discuss the various possible 
origins  for the 
cross-power signal and show that neither local foregrounds, 
nor the known remaining normal galaxies and active galactic nuclei (AGN) can reproduce the measurements.
  These observational results are an important new constraint on theoretical modeling of the near-IR CIB fluctuations.
\end{abstract}


\section{Introduction}

Cosmic backgrounds contain emissions produced during the entire history of the Universe including  from objects individually inaccessible to 
telescopic studies. 
In different spectral regimes, the  cosmic background probes different sources
according to their emission mechanisms. Thus, cosmic X-ray background (CXB, $\sim$[0.5-10] keV) probes both emissions
by accreting black holes (BHs) and thermal X-ray emission from hot ionized gas, such as  in galaxy clusters. Whereas 
the cosmic infrared background (CIB) at the near-IR wavelengths (1-5\um)
is sensitive to stellar emissions \citep[see review by][]{k2005}. 
Correlations between structure in the IR and X-ray backgrounds could arise in two ways:
they could be caused by one or more classes of sources that emit at both IR and X-ray wavelengths; or
they could arise from separate classes of IR-emitting and X-ray-emitting sources that are found
in association on large spatial scales.

At the near-IR, the Galactic and Solar System foregrounds are substantial and, hence, must be known to great
accuracy when estimating the mean levels of the CIB. Thus \citet{k96a,k96b} and \citet{ko} pioneered the measurements of the CIB fluctuations,
 which circumvent many of the difficulties with the foreground subtraction. Indeed, the power spectrum of the
 CIB fluctuations should reflect the clustering of the sources producing them. As the foreground galaxies get eliminated
 to fainter limits, the remaining source-subtracted CIB fluctuations would contain progressively larger fractions
 of the  faint sources inaccessible to current  telescopic measurements.
 A particularly important class here are the sources associated with first stars epoch as the Universe gradually emerged from the ``Dark Ages''.

Current models predict the emergence of the first collapsed objects at redshifts $z\la 30$ (see the review by Bromm \& Yoshida 2011).
The expectation is that at these early times,  a population of black holes (BH) appeared,  
either formed by the deaths of the first stars in a top-heavy initial mass function (IMF), or by monolithic collapse of the
primordial clouds. Although the first luminous objects and galaxies are too faint to observe on their own, it has been proposed that fluctuations in the intensity
 of the cosmic infrared background (CIB) reflect the distribution of these early objects after foreground sources are removed to
 sufficiently faint levels \citep[e.g. see review by][and references cited therein]{k2005}. It has been suggested that these populations may have 
 left a measurable signal in the mean CIB \citep{santos,ferrara} and its fluctuations \citep{kagmm,cooray}. There are
intuitive reasons why CIB anisotropies from the early populations would be measurable: 1)
first stars (and/or the associated BHs) emitted a factor $\sim 10^5$ more luminosity
per unit mass than the present-day stellar populations, 2) their
relative fluctuations would be larger because they span a
relatively narrow time-span in the evolution of the Universe, and
3) they formed at the high peaks of the underlying
density field which amplified their clustering properties.

Intriguingly, there is now a substantial body of  evidence suggesting that the source-subtracted CIB fluctuations, 
discovered in recent {\it Spitzer-}based \citep{kamm1,kamm2} and {\it Akari}-based \citep{akari} studies, may 
arise from new populations which existed in the early Universe. The residual CIB fluctuations 
remain after removing galaxies to very faint levels and arise from populations with a significant clustering
component, but only low levels of the shot noise \citep{kamm3}. This clustering signal exceeds, 
by a large and scale-dependent factor, the fluctuations produced
by the remaining galaxies \citep{kamm1,kari}. As suggested by \citet{kamm1,kamm3} these CIB fluctuations may originate in early populations.
 This found further support in a study by \citet{kamm4} showing that there are no correlations between the source-subtracted IRAC maps and
 the faintest resolved sources observed with the {\it HST} ACS at optical wavelengths, which likely points to the high-$z$ origin of the fluctuations,
 or at least to a very faint population not yet observed by other means. The high-$z$ interpretation of the detected CIB anisotropies has received
  further confirmation in the recent {\it Akari} data analysis which measured source-subtracted CIB fluctuations to wavelengths as short as 2.4 \um\ and
 pointed out that the colors of the fluctuations require their being produced by highly redshifted very luminous sources \citep{akari}. In a
  new step toward understanding the nature of these new populations,  \citet{k12} used {\it Spitzer} data from the SEDS
 program \citep{seds} and for the first time measured the source-subtracted CIB fluctuations up  to $\sim 1^\circ$ showing that the
 amplitude of the CIB fluctuations continues to grow with the scale to more than 10 times that of known galaxies. The data indicate that these fluctuations
 are produced by very faint sources and their angular spectrum is in agreement with an origin in early populations spatially distributed
 according to the standard cosmological model at epochs coinciding with the first stars era.

Such measurements alone, however, do
not provide direct information on whether the emissions in these new populations arise from stellar nucleosynthesis or BH accretion.
If the sources producing these CIB fluctuations contained BHs in sufficient numbers, the latter sub-population would have contributed to the
 CIB fluctuations levels via accretion processes around the BHs. BH accretion also produces
 a large fraction of emission in X-rays which could also produce a potentially identifiable component to the CXB. 
 If the measured CIB fluctuations originate even partly from populations containing a sufficient abundance of BH, then
 the CXB component produced by them should correlate with the CIB providing a way to detect the BH population.

Recent observations with {\em Chandra} \citep{lehmer} resolved $\sim$80-90$\%$ of the [0.5-7] keV energy band CXB into point sources.
The majority of the sources contributing to CXB are AGN powered by accretion onto super massive black holes (SMBH).
However, below the fluxes reached in deep {\em Chandra} observations, most sources are {\it normal} galaxies
whose X-ray emission is largely produced by X-ray binary stars. \citet{cap12} have shown, through angular fluctuation analysis,
 that about 50\% of the unresolved CXB is produced by galaxy groups with the remaining produced by galaxies and AGN.
It was also suggested that, if the large scale excess power observed in the CIB is 
created by the primordial BH at $z>7.5$, then up to 1/3 of the large scale CXB fluctuations could
be produced by them without exceeding the observed power spectrum, while accounting only for a relatively small fraction ($\ll5\%$) of the
total CXB flux. Since high-$z$ sources are expected to be highly biased,
their fluctuation may be detectable despite a smaller contribution to the total CXB flux.

Here we report the first direct evidence of substantial X-ray emission associated with the sources
of the CIB anisotropies uncovered in deep {\it Spitzer}/IRAC data \citep{kamm1,kamm2,k12} and 
briefly discuss the contributions to this signal from the various cosmological candidates.
 This result provides a major clue
to the nature and epochs of the populations producing the source-subtracted CIB fluctuations. We detect 
 correlations which indicate that at least $15 - 20\%$ of the CIB is produced by objects with powerful  X-ray emission.
This proportion is much greater than among the known galaxy populations in the recent Universe. 
If the sources producing the sources-subtracted CIB signal are at high $z$, 
 these findings may suggest a necessity to revise reionization analysis to 
 include substantial contribution from X-ray emissions to the reionization of the Universe.
These observational results also suggest serious revisions in theoretical modeling of the near-IR CIB fluctuations from early
times (cf. Cooray et al 2012a, Yue et al 2013).

 This paper is structured as follows: Sec. \ref{sec:data} discusses the data assembly for the EGS/AEGIS field observed
 by both {\it Spitzer}/IRAC and {\it Chandra}. Sec. \ref{sec:analysis} presents the results of the cross-power analysis,
 identifying a highly statistically significant cross-power between the source-subtracted CIB and CXB. Finally,
 in Sec. \ref{sec:interpretation} we discuss the various possible low- and high-$z$ contributors to the measurements.

\section{Data assembly}
\label{sec:data}

\subsection{X-ray data}
\label{sec:data_x}

The primary X-ray data set used here is the deep
{\em Chandra} ACIS-I AEGIS-XD survey \citep{goulding}
in the area overlapping with the SEDS IRAC survey in
the EGS field. The relevant parameters are listed in Table \ref{tab1}.
The field is located at Celestial / Ecliptic / Galactic
coordinates of (214.91$\arcdeg$, 52.43$\arcdeg$), (180.56$\arcdeg$, 60.00$\arcdeg$), (95.95$\arcdeg$, 59.81$\arcdeg$) and covers
approximately 0.1 deg$^2$.

The {\em Chandra X-ray Observatory} has a peak effective area of 700 cm$^2$ at $\sim$1.2 keV and
superb on-axis angular resolution of $\sim$0.5$\arcsec$ \citep{weiss}.
For imaging surveys, the X-ray telescope is generally coupled with a 16$\arcmin\times$16$\arcmin$ CCD array, ACIS-I
with an average energy resolution of $\sim$130 eV.
The sensitivity window of {\em Chandra} covers the $\sim[0.5-7]$ keV band, and since the CCD records
the energy of the events it is possible to derive multi-band images with a single exposure.

The AEGIS-XD program consists of a series of 66 pointings in the central area of the EGS field.
For our purposes, we employed the 45$\arcmin\times$8$\arcmin$ region that overlaps with the {\em Spitzer}
EGS-SEDS field. Note that this area corresponds to the deepest part of the whole $1\arcdeg \times 16\arcmin$ X-ray survey area.
For every pointing we used level-3 data produced for the {\em Chandra} source catalog,
with the most recent calibration database. Only observations taken in VFAINT mode were considered.
The data have been cleaned of spurious events such 
 cosmic rays as well as instrumental artifacts. Time intervals with high particle background levels have been removed.
A detailed description of the data reduction can be found in \citet{evans}. 
Events have been sorted in arrival time to create odd- and even- listed event files, hereafter A and B subsets. 
The reason for splitting the events in two subsets is explained in the next section. 
In every observation and for each A,B subset, images have been created in the [0.5-2] keV, [2-4.5] keV  and [4.5-7] keV energy bands, respectively.
The choice of this set of bands allows  us to have the same number of counts  ($\sim$1.3$\times$10$^5$ cts ) 
and therefore the same statistical sampling in the three bands.
In the same bands the exposure maps were computed at effective energies of 1.2, 3.2 and 5.5 keV, respectively.
Both images and exposure maps have been rebinned to match the IRAC
maps at 1.2$\arcsec$/pix. 
Finally, for each band and for each subset, all the images and exposure maps have been summed to produce
 mosaic maps. The raw 0.5-2 keV $A+B$ count rate map and the exposure map are shown in 
 the top and central panel of Fig. \ref{fig:maps}, respectively. 
 The count rate map has been smoothed with a gaussian filter of 3.6$\arcsec$ (3 pixels) width to highlight
 features in the image.  
 The mean, cleaned exposure is 640 ksec. Since we are interested in the source-subtracted CXB,
an important step in the data analysis is the removal of point-like and extended sources. 
Thus, in order to remove as many sources as possible we performed a standard source detection 
in the [0.5-2] keV band and a combined [0.5-7] keV band by using the
CIAO tool {\it wavdetect} with a threshold of 10$^{-5}$, corresponding to $<5$ spurious
detections over the whole field of view. 
As a result we detected  303 unique point sources down to fluxes of 7.0$\times$10$^{-17}$ \flux, 
and 1.1$\times$10$^{-16}$ \flux, in the two bands, respectively. 
(No other sources than those detected in these two bands
would  have been  detected in the [2-4.5] keV and [4.5-7] keV energy bands.)
However the flux limit is not constant across the field of view since, as one can notice
from Fig. \ref{fig:maps}, the exposure varies according to the pattern of the tiled observations. Moreover 
the PSF size varies across the field of view. This effect introduces  further 
inhomogeneities in the flux limits. A detailed description of the 
flux limit versus sky coverage is beyond the scope of this paper and can be   
found in \citet{goulding}.
Note that the actual flux limits are dependent on the spectra of the sources. 
Here we assumed that the sources have a typical power-law
spectral index of $\Gamma$=2. In which case, the derived flux limits
can vary by  5\%, 10\%, and 15\% if $\Gamma$ changes by $\pm{0.3}$ in the [0.5-2], [2-4.5] and [4.5-7] keV energy bands, respectively.
The actual CXB flux produced by detected sources
is of the order 1.1 and 2.5$\times$10$^{-8}$ \cgs in the
[0.5-2] and [0.5-7] keV bands, respectively. These values carry an additional $20\%$ uncertainty because
of the spectral model dependence. Since our data are flux-limited in a position-dependent way, the values
stated here are the average value of CXB resolved into point sources across the field of view.
Our brightest source has a [0.5-2] keV flux of the order of 5-6$\times$10$^{-14}$ \flux, which is slightly above
the knee of the Log($N$)-Log($S$) distribution \citep{cap09}. Thus 
a large fraction of CXB flux is not included in the resolved flux  mentioned  above. Moreover
since the X-ray maps used in this analysis are further masked for IR sources, 
the actual fraction of the CXB resolved in our maps cannot be computed in a straightforward way.

In order to remove the detected sources from the maps,
the software computes the distribution of counts within the source 
cell (i.e. the observed counts) for every source, and, assuming it to be Gaussian, 
masks all the source counts in a circular region within 5$\sigma$ of the centroid.
This method does not rely on the actual tabulated PSF FWHM as function of the 
off-axis angle, which is subject to on-orbit calibration uncertainties,  and allows us to limit
the contribution of the PSF wings to the diffuse CXB to  a fraction $<$5$\times$10$^{-7}$. 
 
\citet{erf}  detected  seven extended sources (identified as galaxy groups) in the sky area 
investigated here  by using a wavelet algorithm on
scales of 32$\arcsec$-64$\arcsec$ combined
 with their optical red sequence  and spectroscopic identification.
 This procedure allowed us to mask
clusters and groups of galaxies down to a mass of $\sim$10$^{13}\Msun$.
The circular regions used here to mask extended sources
enclose the projected $r_{200}$ radius, which ensures a highly
efficient removal of the thermal X-ray photons contained in these groups.
Masses and $r_{200}$ were estimated by \citet{erf} with the  X-ray scaling relations  \citep[see e.g.,][]{pratt}
carefully described and tested by \citet{fin07}. As a result, the masking of X-ray sources leaves $\sim$96\% 
of the pixels useful for the CXB fluctuation analysis.
The X-ray mask has been combined with the IR
mask described below \citep{k12} and is shown in the Figure \ref{fig:maps}.
The combination of the
X-ray and IR mask left $\sim$68\% of the map pixels for fluctuation analysis via the FFT.
The remaining counts are thus the CXB, plus the particle background recorded by the detector.
The particle background has been subtracted by tailoring images taken by ACIS-I in stowed mode.
Basically, ACIS was exposed when stowed outside the focal area. Since the particle background
is  not focused, the stowed image simply contains events due to particles.
Such a background level, however, is not constant in time and thus one must find a recipe
to renormalize the stowed image to match the actual background level in the observations.
\citet{hm06}, showed that regardless of  its amplitude, the particle background has a constant spectrum.
In addition all the counts collected by {\em Chandra} in the [9.5-12] keV band have
a non-astrophysical origin (i.e. they are only particle events). Thus, the simple recipe proposed by \citet{hm06} to compute
the particle background level in each band is to scale the stowed images by the ratio $C_{data}$[9.5-12]/$C_{stow}$[9.5-12],
where $C_{data}$ and $C_{stow}$ are total counts measured in the real images and in the stowed image, respectively.
We have then subtracted the corresponding particle background
image for each pointing.
In addition, in  order to compute the CXB fluctuation maps, we derived for every pointing and for every band,
the mean CXB level map
which is dependent on the off-axis angle because of vignetting.
To do this,
we created a map with a total number of counts equal to that of the real data outside the mask
and distributing them according to the relative value of the exposure map. 
The count, mean-value and exposure maps have been then co-added in order
to produce the final mosaic $C_x$, $\langle\,C_x\rangle$ and $E$ maps.
The final fluctuation image is then $\delta F_x=C_x/E-\langle\,C_x\rangle/E$.
With this method we ensure that features likes stripes, dithering and dead pixels 
are carefully reproduced in the mean-value map and therefore do not affect the 
final  $\delta F_x$ map. 

 We also produced random noise maps drawn from two
subsets of events. The events have been sorted in time
and odd- and even-listed photons have been attributed to
images $A$ and $B$, respectively. These maps have the same
exposure time and have been observed simultaneously
so that effects of source variability are removed. In the same way as for real data, we created
$A$ and $B$ fluctuation maps.
The difference of these
maps does not contain celestial signals or any  stable
instrumental effects. For this reason the $\half(A-B)$ difference maps can be used
to evaluate the random noise in the CXB fluctuations maps.
Actually, the cosmic CXB fluctuation maps, $\delta F_x$, have been produced by averaging the $A$ and $B$
data set, so that the auto- and cross-power spectra were evaluated on the  $\half(A+B)$ maps.
The fluctuation count rate maps
have been transformed into flux maps by applying the energy conversion
factors (ecf) listed in Tab. \ref{tab1} under the assumption that
the average X-ray spectrum of the undetected sources could be represented by a power-law with $\Gamma$=2.
Note that the actual spectrum of the sources contributing to the unresolved X-ray background
 is unknown since it is made by a blend of galaxies, clusters and AGN, and for this reason
we have chosen an average  spectral model
of AGN and X-ray galaxies in the [0.5-7] keV band \citep{rana,gil07}.

\begin{deluxetable}{ccccccc}
\tabletypesize{\scriptsize}
\tablecaption{X-ray maps properties\label{tab1}}
\tablewidth{0in}
\tablehead{
\colhead{Band} &
\colhead{N$_{cts}$\tablenotemark{a}} &
\colhead{N$_{cts}$\tablenotemark{b}} &
\colhead{$\langle$N$_{ph}\rangle$/pix} &
\colhead{F$_{lim}$} &
\colhead{$\langle$CXB$_{res}\rangle$} &
\colhead{ecf}\\
& &  
& &
\colhead{\flux} &
\colhead{$\times$10$^{-8}\cgs$} &
\colhead{$\times$10$^{11}$ erg$^{-1}$ cm$^{2}$}
}
\startdata
0.5-2.0 keV & 233867 &133726  & 0.23 & 7$\times$10$^{-17}$ & 1.10$\pm{0.08}$ & 1.55\\
2.0-4.5 keV & 216776 &137838 & 0.23 &  \nodata &  \nodata  &0.67\\
4.5-7.0 keV & 201856 &134808 & 0.23 & \nodata&   \nodata  &0.27\\
0.5-7.0 keV & 652499  &406432& 0.69 &   1.1$\times$10$^{-16}$   &2.5$\pm{0.19}$   &1.07\\
\enddata
\tablenotetext{a}{X-ray photon counts before masking.}
\tablenotetext{b}{X-ray photon counts after masking.} \\

\end{deluxetable}

\subsection{IRAC-based maps}
\label{sec:data_ir}

\begin{figure*}[t!]
\centering
\includegraphics[angle=90,scale=.7]{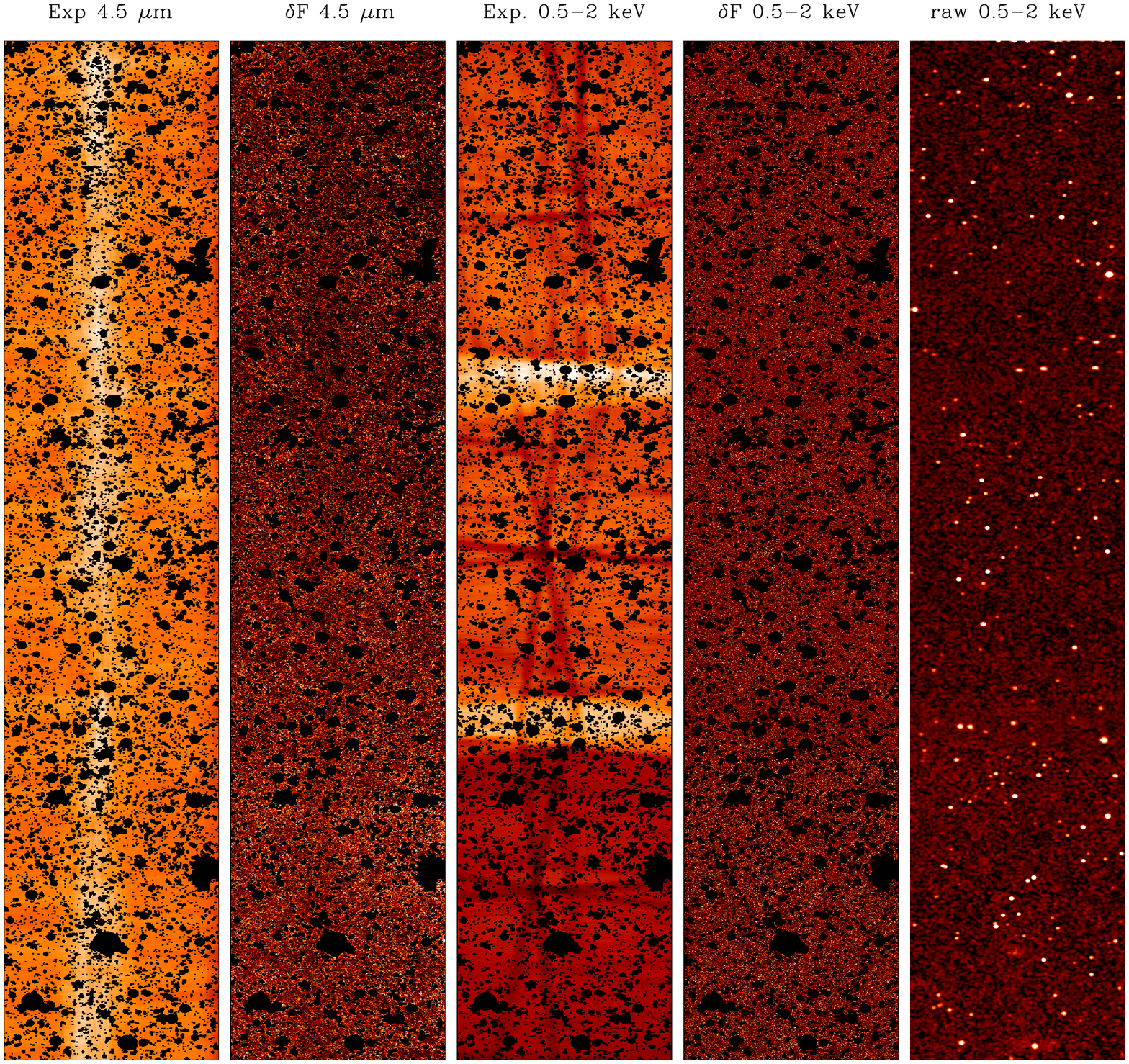}
\caption{From top to bottom : X-ray 0.5-2 keV count-rate map, smoothed with a Gaussian kernel of 3.6$\arcsec$ (3 pixels).
X-ray 0.5-2 keV fluctuation maps in counts rate units.  The X-ray exposure map.  IRAC 4.5 \um\ fluctuation map. 
The IRAC 4.5 \um\ exposure map. In all maps
the black areas represent the mask. The EGS field is located at Celestial / Ecliptic / Galactic
coordinates of (214.91$\arcdeg$, 52.43$\arcdeg$), (180.56$\arcdeg$, 60.00$\arcdeg$), (95.95$\arcdeg$, 59.81$\arcdeg$)
and these sub-images cover $45\arcmin\times 8\arcmin$.} \label{fig:maps}
\end{figure*}

The {\it Spitzer Space Telescope} is a 0.85 m diameter telescope launched into an earth-trailing solar orbit in 2003 \citep{werner2004}.
For nearly 6 years, as
it was cooled by liquid He, its three scientific instruments provided imaging and spectroscopy at wavelengths from 3.6 to 160 $\mu$m.
In the time since the He supply was exhausted, {\it Spitzer} has continued to provide 3.6 and 4.5 $\mu$m imaging with its Infrared Array
Camera (IRAC). IRAC has a $5'\times5'$ field of view, and a pixel scale of $1.2''$, which slightly undersampled the
instrument beam size of $\sim2''$ FWHM \citep{fazio2004}.

The procedure for map assembly is described in our previous papers \citep{kamm1,kamm2}
with an extensive summary, including all the tests, given in \citet{akmm}.
Our IRAC mosaics are prepared from the basic calibrated data (BCD)
product using the least-squares self-calibration procedure described by
Fixsen, Moseley \& Arendt (2000).
The preparation and properties of the IR data obtained in the course of the SEDS program and used
here are discussed in \citet{k12}. The SEDS program was
designed to provide deep imaging at 3.6 and 4.5 \um\ over a total area of about 1 square degree, distributed
over 5 well-studied regions \citep{seds}. The area covered is
 about ten times greater than previous {\it Spitzer} coverage at comparable depth.
While the main use of the SEDS data sets will be the
 investigation of the individually detectable and countable galaxies, the remaining
backgrounds in these data are well-suited for CIB studies, by virtue of their angular scale, sensitivity and observing strategies.
Because of the sufficiently deep coverage with both {\it Chandra} and {\it Spitzer} observations, we have selected the
Extended Groth Strip (EGS; {\it Spitzer} Program ID = 61042) field for this analysis.
The field is located at moderate to high Galactic latitudes to minimize the number of foreground stars
and the brightness of the emission from interstellar medium (cirrus). It also lies at relatively high ecliptic latitudes,
which helps minimize the brightness and temporal change in the zodiacal light from interplanetary dust. The observations were
carried out at three different epochs, spaced 6 months apart. 
At each wavelength, the frames are also processed in several different groups to provide
multiple images that can be used to assess random and systematic errors. The noise is obtained by separating the full sequence of frames into the
alternating even and odd frame numbers. Comparison of these ``A'' and ``B'' subsets,
through construction of $\half($A-B$)$ difference maps,
provides a good diagnostic of the random instrument noise because the A and B subsets only
differ by a mean interval of $\sim100$s. 

We also examined shallower ($\sim3$ hr integration) 5.8 and 8 $\micron$
observations of the EGS field that were obtained during {\it Spitzer's} cryogenic
mission (program ID = 8). However, even with application of the
self-calibration, we find that the resulting images have background problems.
Some of the problems are likely to be intrinsic and related to cirrus, i.e.
thermal emission from interstellar dust. These observations cover a longer strip
of the EGS than the SEDS observations. At the extreme end of the 8 $\micron$
image ($\sim0.5\arcdeg$ from the SEDS region) there is clearly diffuse emission
from cirrus, which is also evident in the {\it IRAS} 100 $\micron$ images and
the LAB HI images (Neugebauer et al. 1984; Kalberla et al. 2005). The
5.8 $\micron$ data show additional background problems that are not correlated
with the 8 $\micron$ data. These problems appear to be related to greater
instability of the detector offset at 5.8 $\micron$, which can be confused with
temporal changes in the zodiacal light. Self-calibrating the 5.8 $\micron$ data
{\it without} the subtraction of the estimated zodiacal light normally applied
by the BCD pipeline provides a better, but still not satisfactory, result. The
background issues at both 5.8 and 8 $\micron$ may be compounded by the observing
strategy. The SEDS strategy stepped across the full length of the field
relatively quickly, and then accumulated depth by repeated observations, while
the cryogenic observations accumulated the full depth of coverage at each
pointing before moving on to another location along the EGS field. Because of
these background issues and the higher noise levels in these data, cross
correlations of 5.8 and 8 $\micron$ emission with X-ray emission did not yield
any significant results to present in this paper.

The region selected for the joint CXB-CIB analysis is about $\sim 8^\prime\times45^\prime$ in size.
The common mask from the IRAC 3.6\um\ and 4.5\um\ bands and the X-ray bands was used, with about $\simeq 32\%$ of
the pixels lost to the analysis. An example of the CIB fluctuation maps is shown in Fig. \ref{fig:maps}.
\section{Fluctuation analysis}
\label{sec:analysis}

\subsection{Definitions}
The maps under study are clipped and masked for the
resolved sources, yielding the fluctuation field, $\delta
F(\vec{x})$. The Fourier transform,
$\Delta(\vec{q})= \int \delta F(\vec{x}) \exp(-i\vec{x}\cdot\vec{q}) d^2x$ is
calculated using the FFT. The power spectrum in a single band $n$ is $P_n(q)=\langle |
\Delta(\vec{q})|^2\rangle$, with the average taken over
all the independent Fourier elements which lie inside the radial interval $[q, q+dq]$. 
Since the flux is a real quantity,
 only one half of the Fourier plane is independent, so that at each $q$ there are $N_q/2$ independent
 measurements of $\Delta(\vec{q})$ out of a full ring with $N_q$ data. 
A typical rms flux fluctuation is $\sqrt{q^2P_n(q)/2\pi}$ on the
angular scale of wavelength $2 \pi/q$. The correlation function,
$C(\theta) = \langle \delta F(\vec{x})\cdot \delta F(\vec{x}+ \vec{\theta})\rangle$, is uniquely related to $P_n(q)$ via Fourier
transformation. If the fraction of masked pixels in the maps is too high, the large-scale map properties
cannot be computed using the Fourier transform and instead the maps must be analyzed by
direct calculation of $C(\theta)$, which is immune to mask effects. In this study,
the clipped pixels occupy $\simeq 32\%$ of the maps which allows for a robust FFT analysis; this issue has been addressed
in great detail in the context of the {\it Spitzer}-based CIB studies in \cite{kamm1, akmm, k12}.


We characterize the similarity of the fluctuations measured in different bands
via the cross-power spectrum, which is the Fourier transform of the cross-correlation function
$C_{mn} (\theta) =\langle \delta F_m(\vec{x})\cdot \delta F_n(\vec{x}+ \vec{\theta})\rangle$.
The cross-power spectrum is then given by $P_{mn} (q) = \langle \Delta_m(q) \Delta^*_n(q)\rangle = {\cal R}_m(q) {\cal R}_n(q) + {\cal I}_m(q) {\cal I}_n(q)$ with ${\cal R, I}$
standing for the real, imaginary parts. Note the cross-power of real quantities, such as the flux fluctuation, is always real,
but unlike the single (auto-) power spectrum the cross-power spectrum can be both positive and negative.

The errors on the power have been computed by 
using the classical Poissonian estimators so that for 
the auto-power $\sigma_{P_{n}}(q)={P_{n}}(q)/\sqrt{0.5\,N_q}$ and for the cross-power
$\sigma_{P_{mn}}(q)=\sqrt{{P_m}(q){P_n}(q)/N_q}$. These errors have been verified to be accurate to
 better than a few percent from comparison to the intrinsic standard deviation of the Fourier amplitudes at the various $q$.

\subsection{CXB power spectra}

The analysis of the fluctuations of the CXB has been performed in the
{\em Chandra} [0.5-2] keV, [2-4.5] keV and [4.5-7] keV bands.
We evaluated the power spectra and their relative errors from the individual {\em Chandra} masked maps
as well as from the ${\half(A-B)}$ image. The final power spectrum of CXB fluctuations, $P_X$, is therefore
evaluated as $P_{\half(A+B)}-P_{\half(A-B)}$ with correspondingly propagated errors.
The X-ray count maps, however, have an occupation number of $<$1 cts/pix, so
the Gaussian behavior of their variance is not guaranteed especially at small scales.
Correspondingly, we evaluated the mean number of photons per $N_q/2$ elements in the 
Fourier domain. 

\begin{figure}[t]
\centering
\includegraphics[angle=0,height=3in]{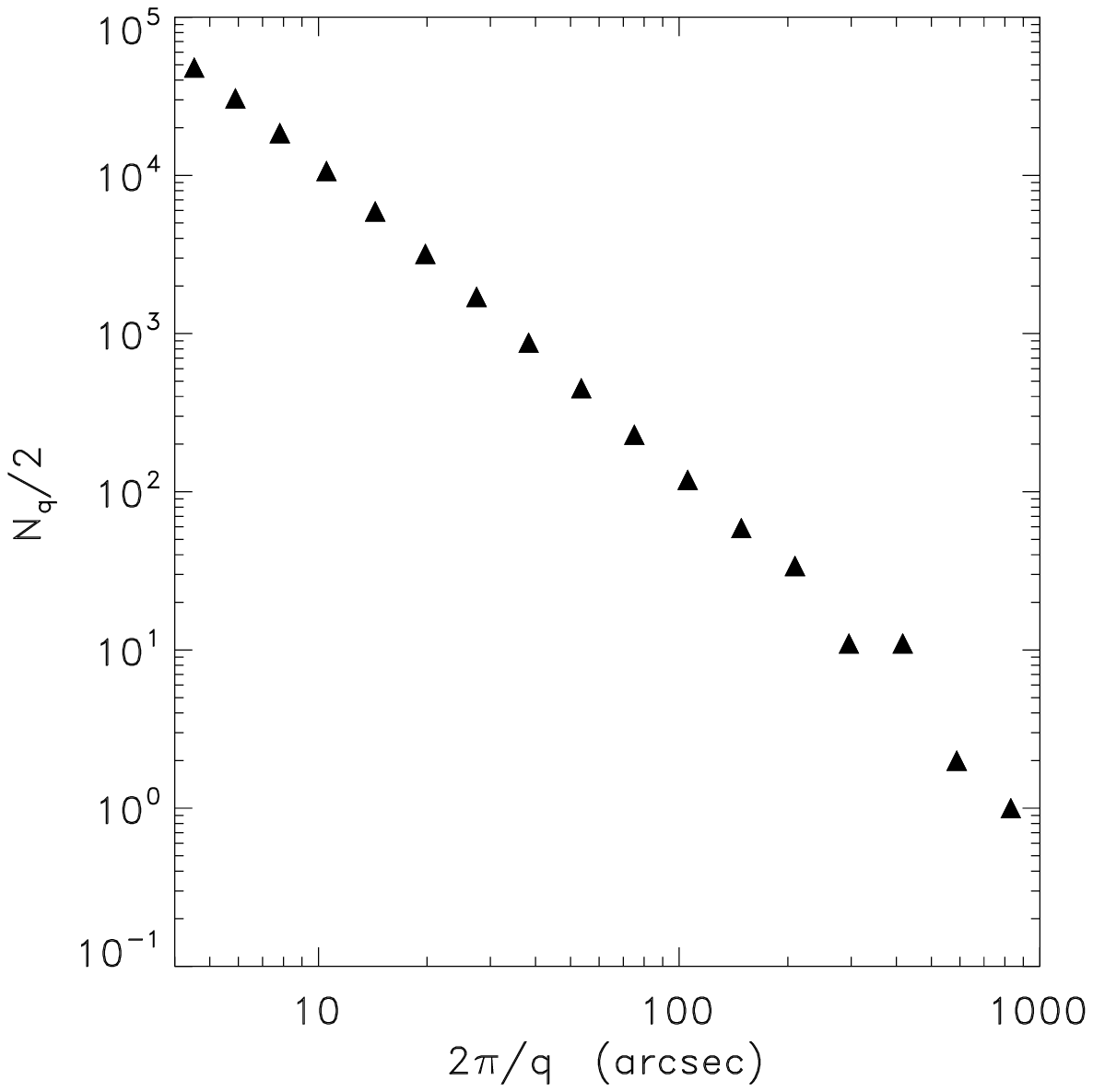}
\includegraphics[angle=0,height=3in]{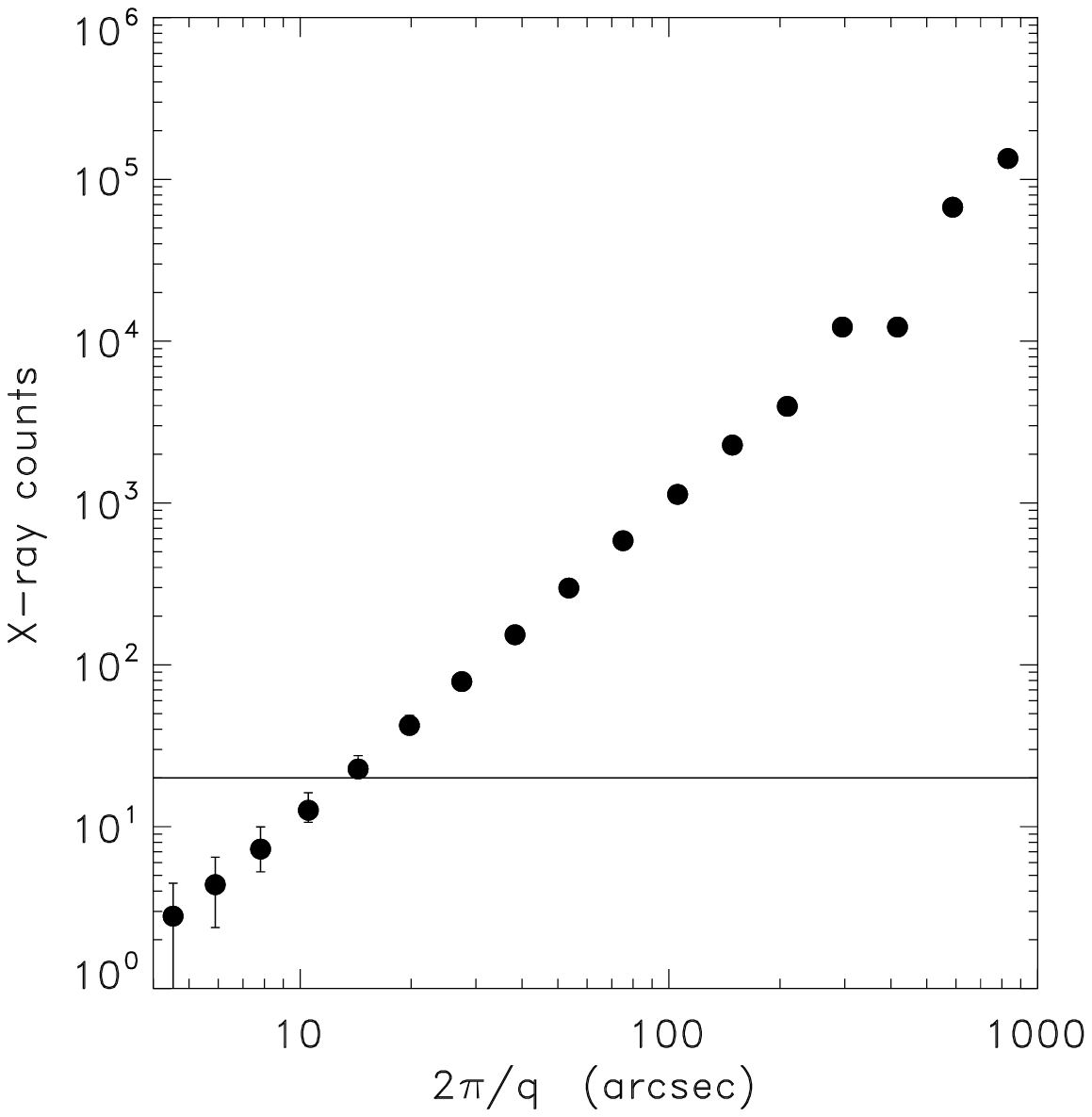}
\caption{\small{Left: Number of independent Fourier elements per bin that went into determining the power spectrum for each field.  Right: The
average number of photon counts per Fourier element adopted for determining the power spectrum as function of angular scale.}
 \label{fig:bin}}
\end{figure}

The left panel of Figure \ref{fig:bin} shows the number of independent Fourier elements, $N_q/2$,
as a function of $2\pi/q$. The right panel shows 
the mean number of  X-ray photons per element  [i.e. $N_{cts}/(N_q/2)$] as function of angular scale, 
where $N_{cts}\approx 135000$ (Tab. \ref{tab1}). 
A limit of 20 cts/element is taken as a practical division between Gaussian and Poissonian regimes. 
The figure shows that below 10$\arcsec$,
X-ray counts are in the Poissonian regime, and therefore we limit our analysis of
 auto- and cross-power spectra to scales $>10\arcsec$
to avoid biases introduced by low-count statistics.

In order to take into account the effects of sensitivity variation across the field
of view, in every pixel the fluctuation field $\delta F_{x}(i)$ has been weighted by a factor $E(i)/\langle E \rangle$ where $E(i)$ is
the effective exposure at the pixel $i$ and $\langle E \rangle$ is the mean exposure in the field.
The clipped and cleaned maps were Fourier transformed and power spectra evaluated. 

\begin{figure}[t]
\centering\centering
\hspace{7cm}
\includegraphics[angle=0,width=1\textwidth]{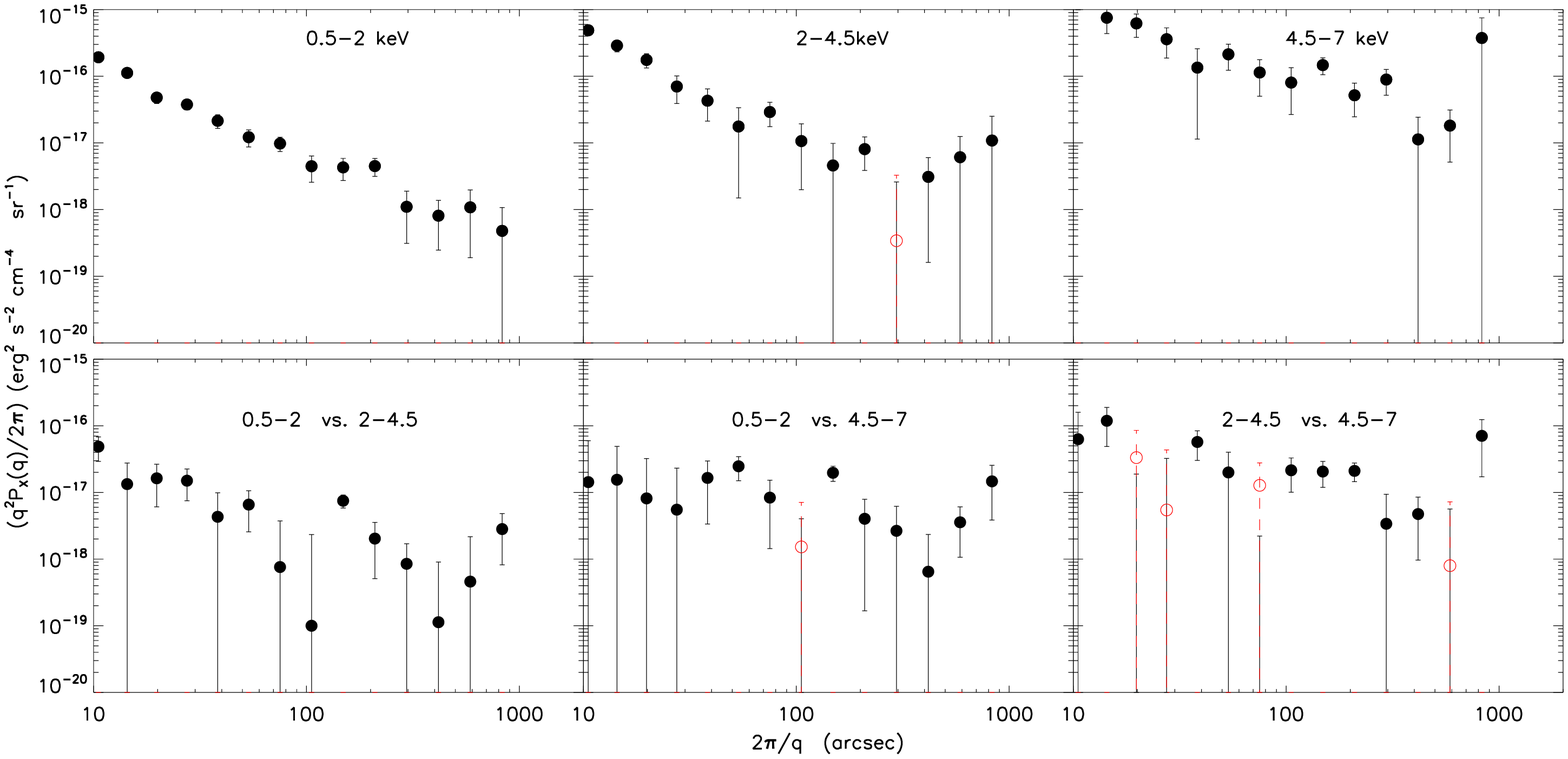}
\caption{Source-subtracted CXB fluctuation power spectra ($top~row$) in the [0.5-2] keV, [2-4.5] keV and [4.5-7] keV energy bands and their relative
 cross-power spectra ($bottom~row$). 
Open red circles and error bars represent the negative power points shown in absolute value for a better highlighting of the signal.}\label{fig:x-power}
\end{figure}

The binning of the power spectrum in angular scale is the same for all the energies sampled here.
The relative sampling error (cosmic variance) on the determined power is $[\half N_q]^{-\half}$, and so the power spectrum
 is not determined highly accurately at the largest angular scales ($\gtrsim250''$) of the EGS field where $\half N_q \lsim 10$.
The X-ray power spectra measured in the three  X-ray energy bands are shown in
Fig. \ref{fig:x-power}. 

The [2-4.5] keV  and [4.5-7] keV power spectra are noisier  than that
measured in the [0.5-2] keV band since in those bands  the particle background
is dominant with respect to CXB.
 In order to probe whether the source-subtracted maps at the different energy bands contain the same
populations, we computed the cross-power spectrum between each pair of maps.  
This analysis shows  that  the cross-power spectra between the hard bands and the [0.5-2] keV band 
generally have lower amplitudes than  the corresponding auto-power spectra, especially on smaller scales.
This suggests that the population of sources producing
the [0.5-2] keV CXB fluctuations can be  substantially different from that producing the  hard X-ray CXB.
Such a conclusion can be confirmed by computing the level of coherence of the signal of every band pairs. 
As in \citet{k12} we can express the common contribution of the sources to {\it both} IR and X-ray signals in terms of
coherence, ${\cal C}(q)=\frac{|P_{\rm m,n}(q)|^2}{P_{\rm m}(q)P_{\rm n}(q)}$.  The coherence can also
 be interpreted as   the fraction of
 the emission due to the common populations so that 
 ${\cal C}\sim\zeta_{m}^2\zeta_n^2$, where $\zeta_{m}$ and $\zeta_m$ are the fractions of the emissions 
produced by the common population in the probed $m$   and $n$ X-ray bands.
As a result, we find that for the band pairs [0.5-2] keV/[2-4.5] keV and [0.5-2] keV/[4.5-7] keV, ${\cal C}\sim$0.1 and ${\cal C}\sim$0.05, respectively.
Thus only 30\% of the [0.5-2] keV emitters contribute also to the [2-4.5] keV power, while $<20\%$ of the 
[0.5-2] keV emitters contribute also to the  [4.5-7] keV power.
The mean  level of coherence between  [2-4.5] keV and [4.5-7] keV is ${\cal C}\sim$0.3-0.4, but with large uncertainties.

It is important to emphasize, in the context of the discussion below (Sect. 4.), that the interpretation of the CXB power and cross-power spectrum carries an intrinsic
source of uncertainty  due to the contribution of the Galaxy.  Thus, the results shown in Fig. \ref{fig:x-power} for individual bands present an {\it upper} 
limit on the unresolved extragalactic CXB fluctuations because they contain the contribution from the Galaxy which is more prominent at the softest energies.
Although our inspection of ROSAT Galaxy diffuse emission maps in this field does not show any well defined structure, 
the actual shape of the Galaxy's diffuse emission power spectrum is unknown on these scales. 
\citet{sli} measured the power spectrum of the ROSAT soft X-ray background fluctuations and showed that its shape and amplitude is a strong function of the Galactic coordinates. 
However, their measurements were obtained on scales larger than $\sim$10$\arcmin$
limiting any direct comparison to the CXB fluctuations in our field. Nevertheless, their Fig. 9 shows that the Galaxy component, at high Galactic latitudes, is approximately white noise at sub-degree scales. 
A more accurate measurement will be possible only with the forthcoming launch of {\it eROSITA} \citep{predehl} in late 2014. 
Thus, while irrelevant for the CXB-CIB cross-power spectrum (see below),  correcting for the 
Galaxy would {\it reduce} our estimate of the extragalactic CXB auto-power spectrum, particularly on the smallest scales. 
Since the Galaxy mostly emits below 1 keV, this could be the reason for
a low-level of cross-correlation between [0.5-2] keV and [2-4.5] keV-[4.5-7] keV maps. 

\subsection{CIB power spectra}
\begin{figure}[t]
\centering
\hspace{2in}
\includegraphics[angle=0,width=1\textwidth]{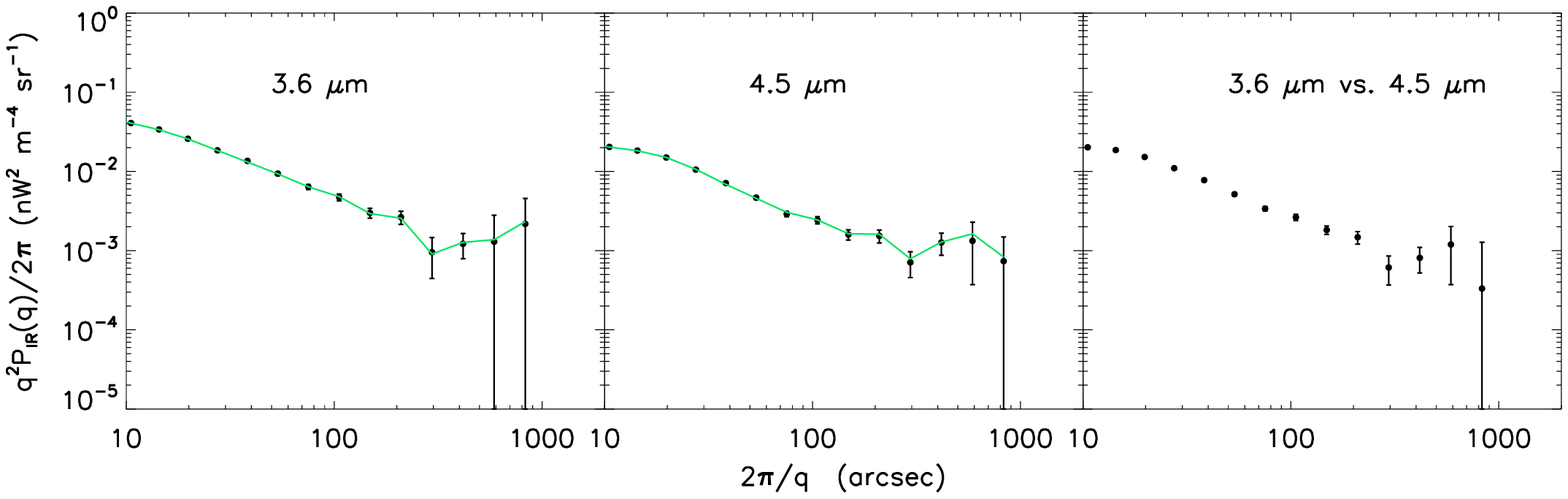}

\caption{Left and central panels: The 3.6 $\mu$m and 4.5 $\mu$m CIB fluctuations power spectra, 
respectively in the EGS field. Right: 3.6 $\mu$m vs 4.5 $\mu$m fluctuations
cross-power spectrum in the EGS field. Green lines show CIB fluctuations 
evaluated after applying {\it only} the IR mask (instead of the IR+X-ray mask) to the data as in \citet{k12}. } \label{fig:irps}
\end{figure}

 In Fig. \ref{fig:irps} we show the auto-power spectra of the IRAC 3.6\um\ and 4.5\um\ maps and their cross-power power spectrum.
 The CIB fluctuation spectra evaluated in this work are in excellent agreement with those
derived by \citet{k12} in the original EGS field even with the additional masking of X-ray detected sources. 
Power spectra with or without the additional X-ray masking agree to better than $5\%$ on all scales,
 as shown by solid symbols and green lines.
 This is consistent with the populations responsible for the CIB fluctuation signal being unrelated
 to the remaining known galaxy or galaxy cluster populations in the field.

\subsection{CIB-CXB cross-power spectra}

\begin{figure}[h!]
\centering
\includegraphics[width=\textwidth,angle=0]{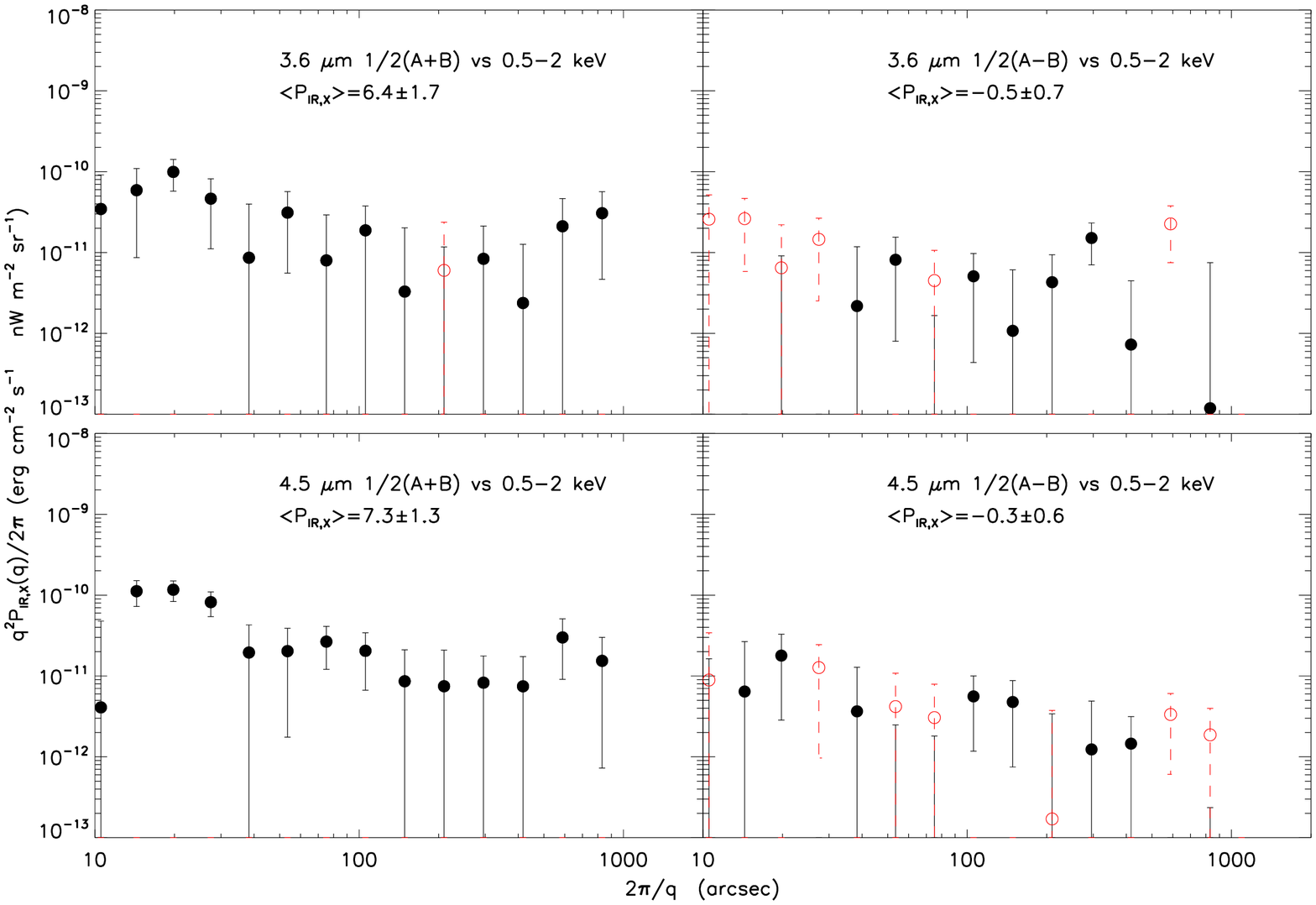}

\caption{(Top left) The fluctuations cross-power spectrum between IRAC 3.6 \um\ $\frac{1}{2}$($A+B$) and {\em Chandra} [0.5-2] keV. 
(Top Right) The fluctuations cross-power spectrum between IRAC 3.6 \um\ $\frac{1}{2}$($A-B$) and {\em Chandra} [0.5-2] keV.
The bottom row shows the same, but for IRAC 4.5 \um\ and {\em Chandra} [0.5-2] keV. The labels in the plots
list the average cross-power measured on the angular range 10$\arcsec$-1000$\arcsec$ in units of 10$^{-20}$ \flux nW m$^{-2}$ sr$^{-2}$.
Open red circles and dashed red error bars represent the absolute values of negative power points.} \label{fig:crosspower}
\end{figure}

\begin{figure}[h!]
\centering
\includegraphics[width=\textwidth,angle=0]{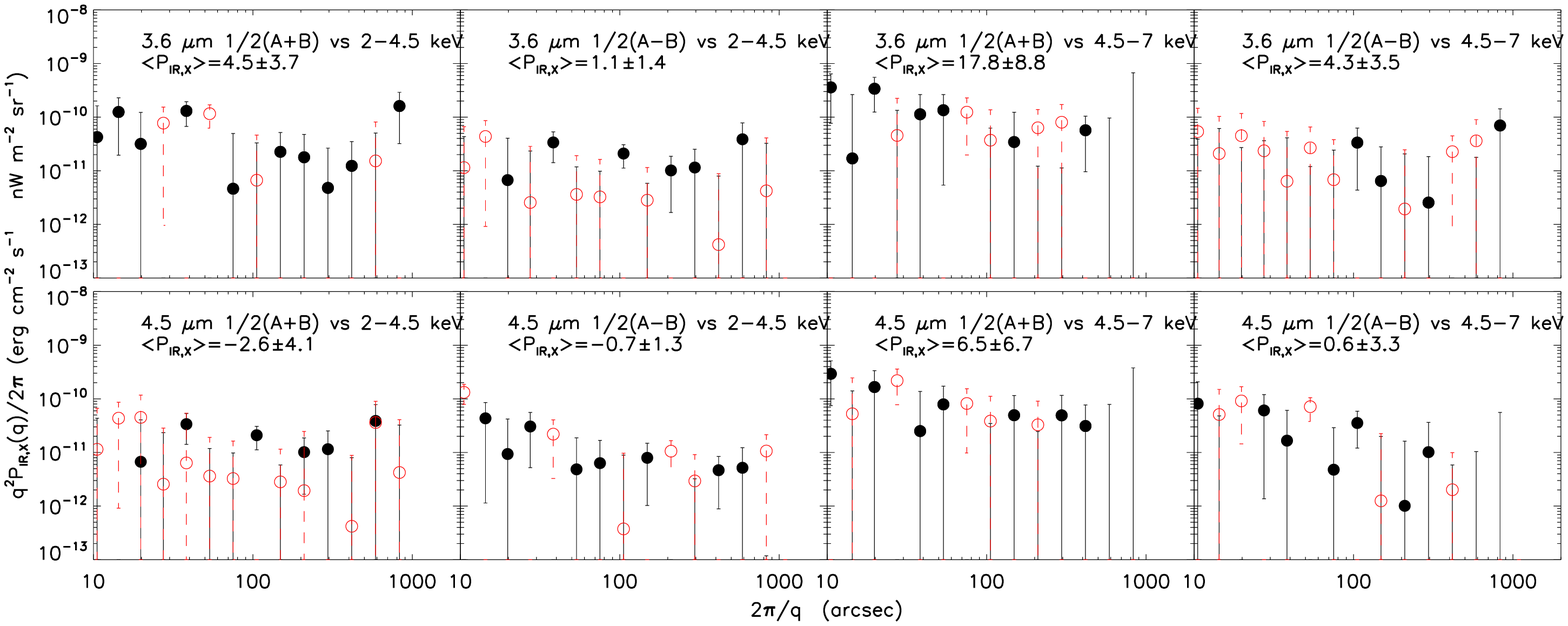}

\caption{
Same as Fig. \ref{fig:crosspower} but for the [2-4.5] keV band (left panels) 
and [4.5-7] keV band (right panels). Note that in these hard X-ray bands
the measured cross power is consistent with zero in all cases. 
\label{nopower}}
\end{figure}

In order to establish if the fluctuations in the source-subtracted CXB and CIB maps
have been totally or partly produced by a population of sources sharing  the same environment (or even being the same sources), 
we performed the cross-power analysis and evaluated $P_{\rm IR, X}(q)$. Since the X-ray and
 IR noise are uncorrelated, the cross-power of the instrument noise contributions should alternate around zero.
  The cross power-spectrum between
IRAC 3.6\um\ and 4.5\um\  source-subtracted CIB fluctuations and
 {\em Chandra} [0.5-2] keV
fluctuations are shown in Fig. \ref{fig:crosspower}, where we find a statistically significant cross-power.
The same is plotted for IRAC 3.6\um\ and 4.5\um\ versus  {\em Chandra} [2-4.5] keV and  {\em Chandra} [4.5-7] keV,
in Fig. \ref{nopower}, where we do not find statistically significant detection.

\begin{figure*}[h!]
\centering
\includegraphics[width=3.in]{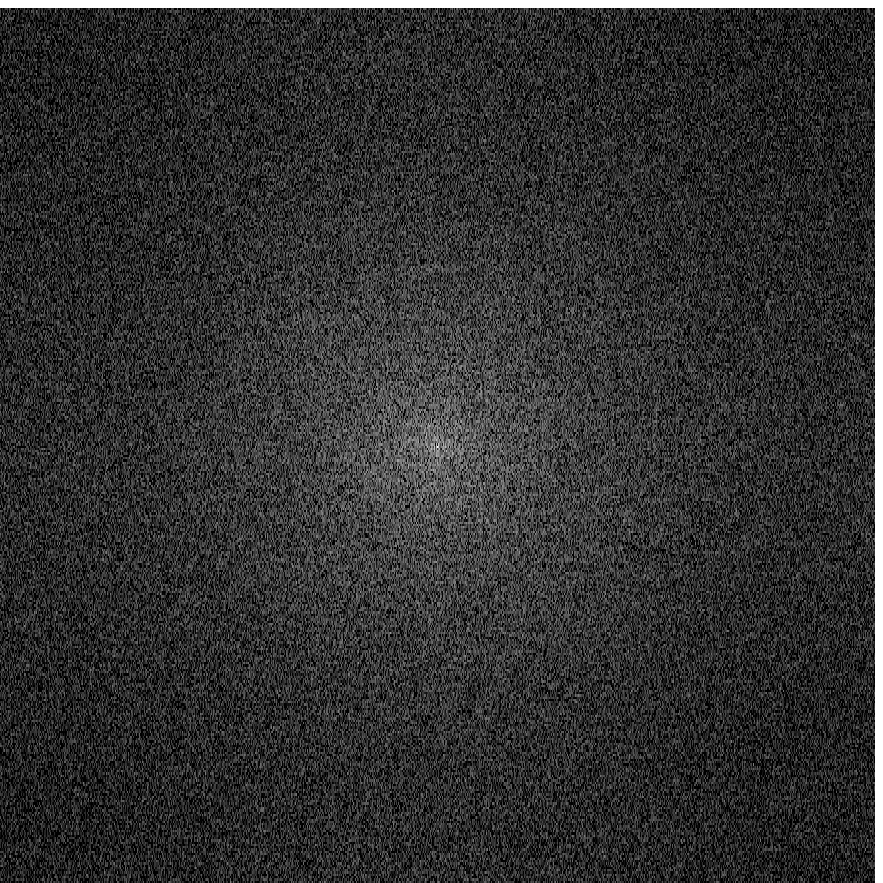}~~
\includegraphics[width=3.in]{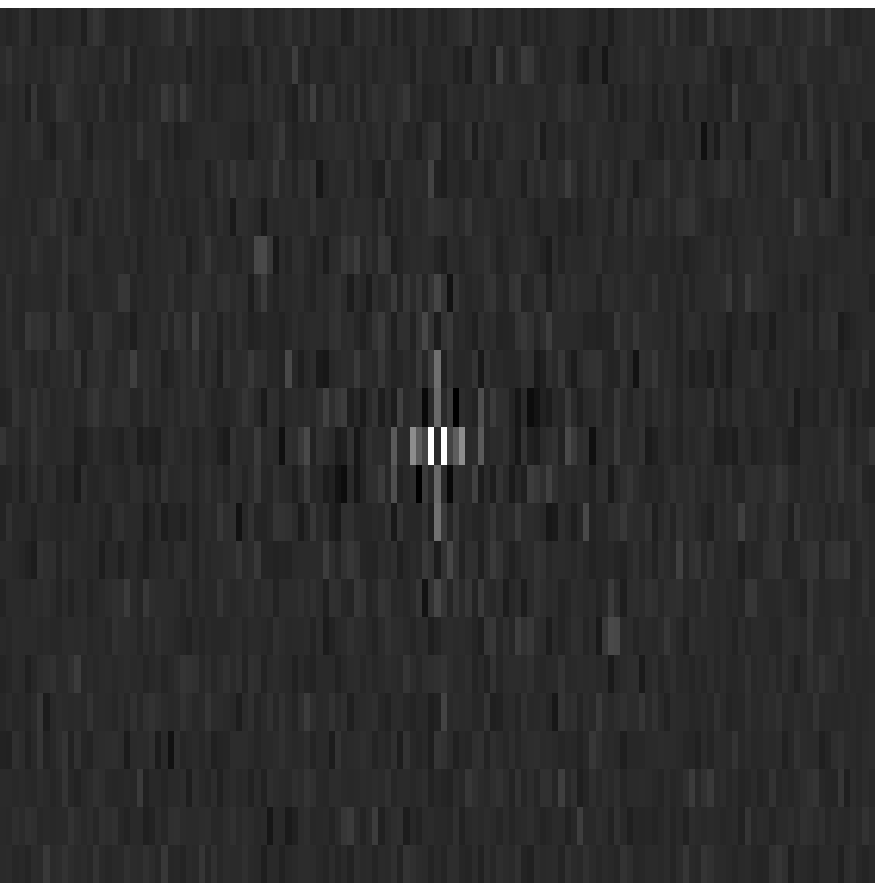}
\caption{ The unbinned 2-dimensional $|P_{\rm IR, X}(q)|$ for 4.5 \um~vs [0.5-2] keV is shown in the left panel with 
a logarithmic scale. The frequency space spans $-2\pi/2.4'' < q < 2\pi/2.4''$ on each axis. 
The right panel zooms in on the low-frequency 
(large-scale) part of $P_{\rm IR, X}(q)$ for $-2\pi/38.4'' < q < 2\pi/38.4''$, shown on a linear scale. Here the 
frequency space pixels are clearly not square because of the rectangular shape of the original image.
The figures show that at both small and large spatial scales, there are not evident artifacts (e.g. asymmetries or large outliers) 
in the 2-D cross-power that would affect the binned 1-D result (Fig. \ref{fig:crosspower}, lower left).
\label{fig:fft}}
\end{figure*}

We evaluated the overall significance of the cross-power by averaging the results over the whole angular range  ($10'' < 2\pi/q < 1000''$), 
computing the mean and its standard deviation. We also evaluated the significance
from the actual dispersion of the un-binned data and found identical results.
In Fig. \ref{fig:fft} we display the full 2-dimensional cross power spectrum, $P_{\rm IR,X}(\vec{q})$, for 4.5 $\mu$m and [0.5-2] keV. 
The  mean power-spectra for every band pair investigated here are reported in Tab. \ref{tab2}.
We find mean correlations at $\sim$3.8$\sigma$ and 5.6$\sigma$ significance 
between the [0.5-2] keV band and IRAC  3.6 and 4.5 $\micron$ bands respectively. 
Stripe-type artifacts and gradients in the images are mapped onto axes in the Fourier representation. So 
although the Fourier maps look reasonably clean, we also evaluated the CIB vs. CXB cross-power
after masking the axes in the Fourier domain. The results are consistent within 1$\sigma$ with those reported in Tab. \ref{tab2}, although 
less significant because of the reduced number of data points introduced by such a masking. 

\begin{figure*}[h!]
\centering
\includegraphics[width=0.32\textwidth,angle=0]{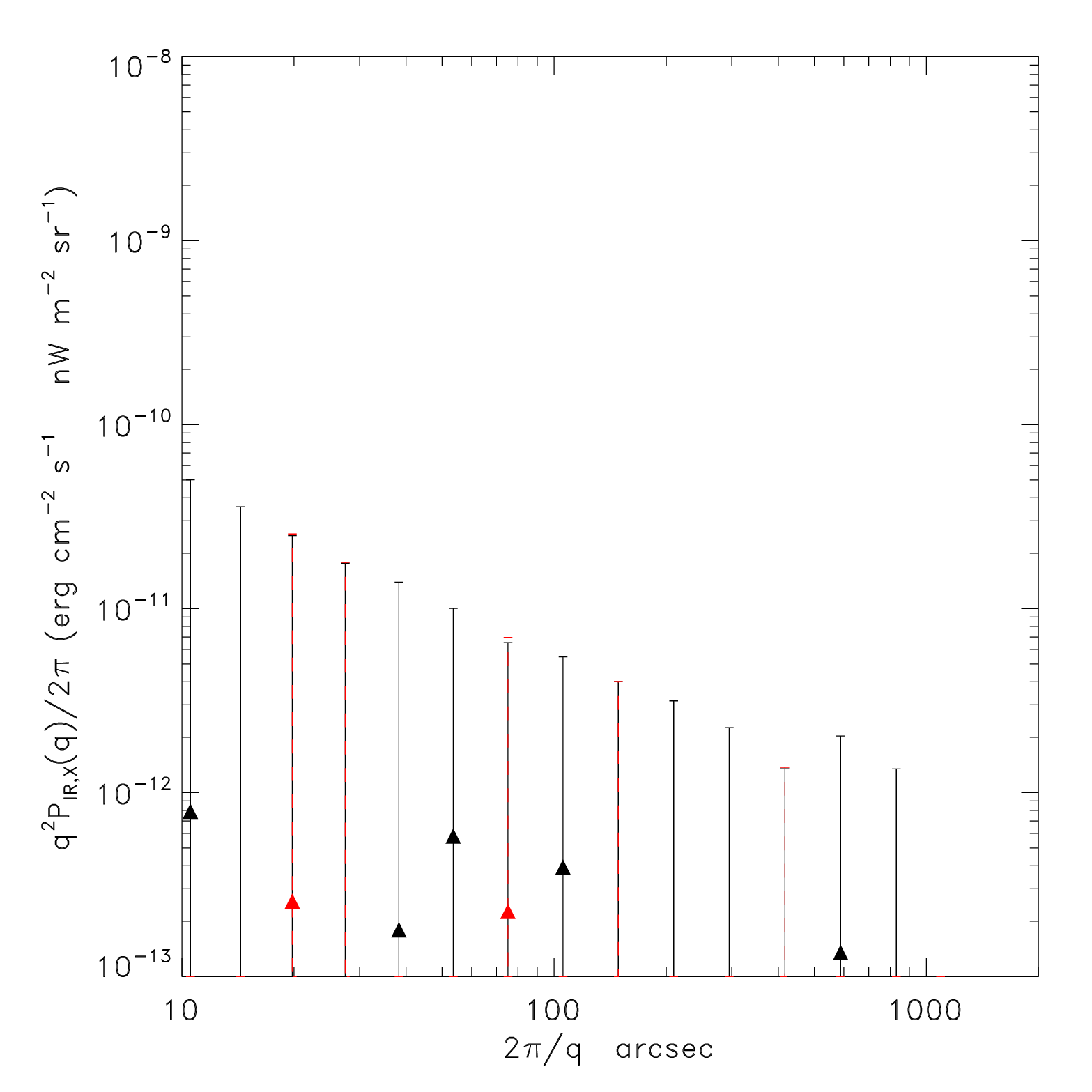}
\includegraphics[width=0.32\textwidth,angle=0]{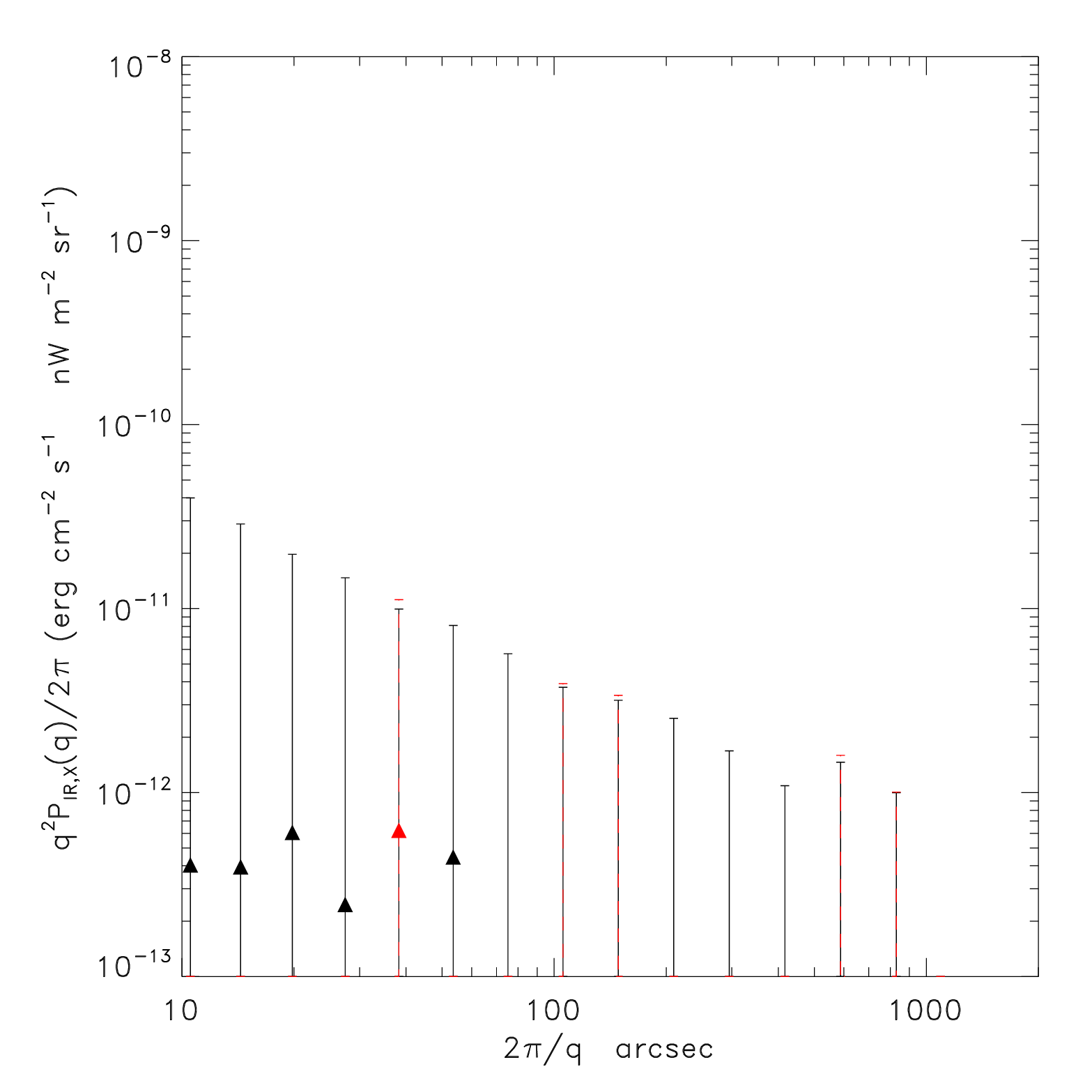}
\includegraphics[width=0.32\textwidth,angle=0]{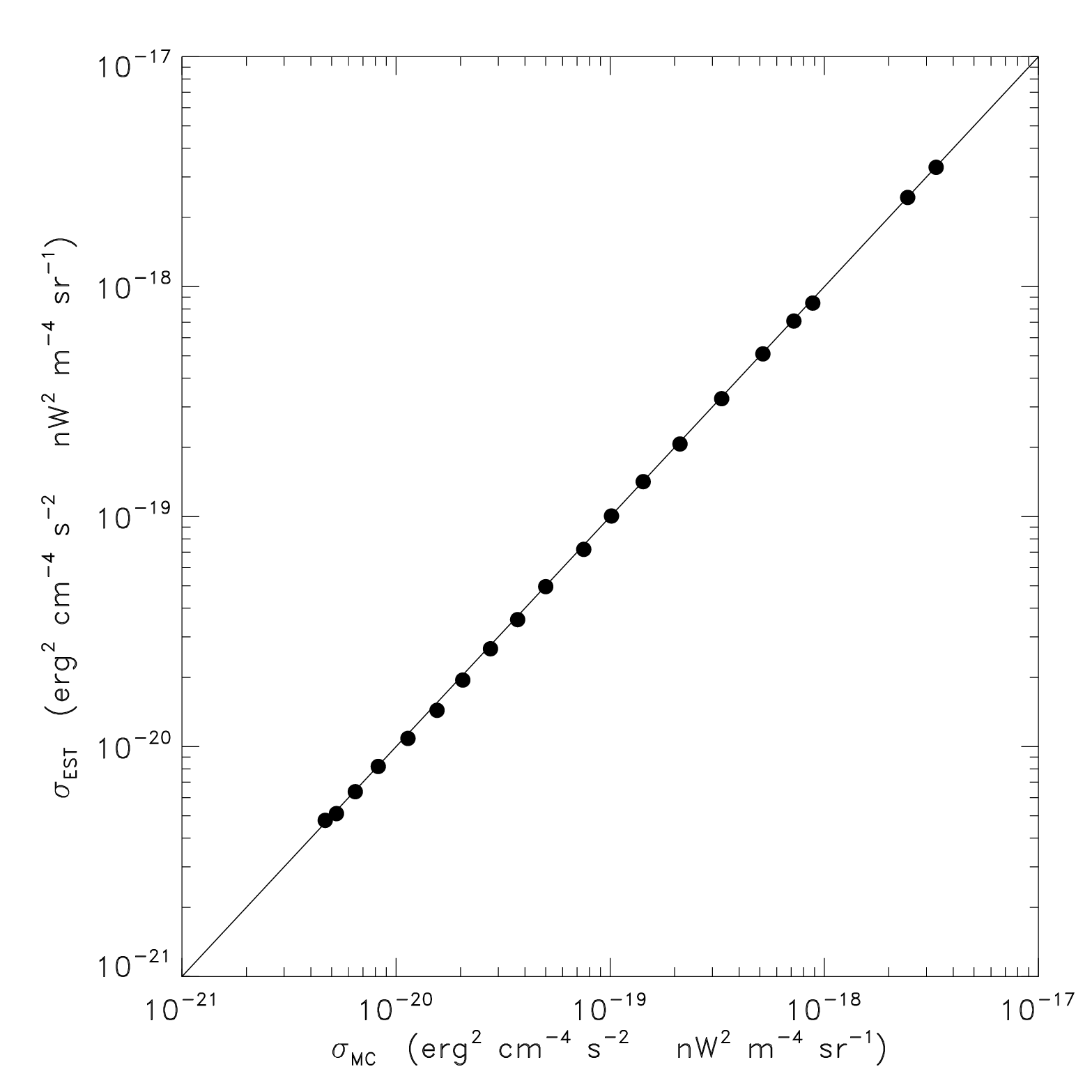}
\caption{\label{fig:mc}  Left panel: The mean [0.5-2] keV vs 3.6 $\mu$m cross power obtained with 1000 Monte Carlo simulations and its standard deviation;
the absolute values of negative cross-powers are plotted in red. The same is plotted in the central panel but referred to [0.5-2] keV vs 4.5 $\mu$m cross power. 
In the right panel we show the relation between errors measured with the actual dispersion of the Monte Carlo realizations ($x$-axis) and that
measured with our estimator ($y$-axis).}
\end{figure*}

The [2-4.5]  keV  and [4.5-7] keV bands do not show significant  cross-correlation with IRAC band as shown in Tab. \ref{tab2}.
We tested if the observed cross-correlation could have been produced by  spurious 
instrumental features by cross-correlating the X-ray $\half($A+B$)$ maps with $\half($A-B$)$ IR maps and  computed their average 
cross-power for every X-ray and IR band pair. 
The results of this analysis are plotted in 
Figs. \ref{fig:crosspower} and  \ref{nopower} and listed
in Tab. \ref{tab2}. We note that  these cross-power spectra are 
always consistent with zero and, as far as 3.6 $\mu$m and 4.5 $\mu$m versus
[0.5-2] keV bands are concerned, the detected signal in the data  is much larger than the cross-correlation between X-ray and IR noise maps.
This cross correlation  provides 
an estimate of the noise contribution for our analysis and a probe for systematic spurious power in the data.  
\begin{deluxetable}{ccccccccc}
\tabletypesize{\scriptsize}
\tablewidth{0in}
\tablecaption{Mean P$_{IR,X}$ in units of 10$^{-20}$\flux nW m$^{-2}$ sr$^{-1}$ computed over the 
[10$\arcsec$-1000$\arcsec$] angular range.\label{tab2} }
\tablehead{
\colhead{Bands} &
\multicolumn{2}{c}{0.5--2 keV}  & &
\multicolumn{2}{c}{2--4.5 keV}  & &
\multicolumn{2}{c}{4.5--7 keV}\\
\colhead{$E_{\rm eff}$} &
\multicolumn{2}{c}{1.2 keV} & &
\multicolumn{2}{c}{3.2 keV} & &
\multicolumn{2}{c}{2.3 keV}\\
\cline{2-3} \cline{5-6} \cline{8-9}
\colhead{} &
\colhead{ $\langle P_{IR,X}\rangle$ }&
\colhead{ $\langle P_{\half A-B,X}\rangle$ }& &
\colhead{ $\langle P_{IR,X}\rangle$}&
\colhead{$\langle P_{\half A-B,X}\rangle$} &  &
\colhead{$\langle P_{IR,X}\rangle$}&
\colhead{$\langle P_{\half A-B,X}\rangle$}
}
\startdata
3.6$\micron$ & {\bf 6.4$\pm{{\bf 1.7}}$}  & -0.5$\pm{0.7}$ & & 4.5$\pm{3.7}$ & 1.1$\pm{1.4}$& & 17.8$\pm{8.8}$  & 4.3$\pm{3.5}$ \\
4.5$\micron$  &{\bf  7.3$\pm{{\bf 1.3}}$}  &  -0.3$\pm{0.6}$ & &  -2.6$\pm{4.1}$ & -0.7$\pm{1.4}$ &  & 6.5$\pm{6.7}$  & 0.6$\pm{3.3}$
\enddata
\tablecomments{Bold text indicates the statistically significant results.}
\end{deluxetable}
We also have cross correlated the CIB fluctuations with our 
particle background model. For the 3.6 $\mu$m and 4.5 $\mu$m
vs. [0.5-2] keV band pairs, the cross power spectra have amplitudes 
of ($-2.7\pm{2.2})\times$10$^{-20}$\flux nW m$^{-2}$ sr$^{-1}$
and (0.7$\pm{1.6})\times$10$^{-20}$\flux nW m$^{-2}$ sr$^{-1}$ 
respectively and thus cannot account for the observed signal. 
To further check the robustness of our results we also 
calculated cross-power spectra of our
 X-ray images with 1,000 random CIB fluctuation maps
constructed by resampling the original masked maps.
In Fig. \ref{fig:mc} we show that, at every scale, no statistically significant cross-power signal can be  recorded.
Moreover such a  test confirms that the amplitude of the estimated errors 
are consistent with the errors obtained by measuring the dispersion of the 
power in Fourier space. To illustrate this we show, in the right panel of Fig. \ref{fig:mc}, that the errors estimated
from the dispersion of the measurement in the Monte Carlo simulation are equivalent to those derived from our estimates.
In a final test we divided the field in two equal parts (left and right sides in Fig. 1) and recomputed the cross-power
between [0.5-2] keV and 3.6 $\mu$m and 4.5 $\mu$m. On the left side we obtain P$_{IR,X}$=(5.7$\pm{2.3})\times$10$^{-20}$\flux nW m$^{-2}$ sr$^{-1}$
and (8.2$\pm{1.8})\times$10$^{-20}$\flux nW m$^{-2}$ sr$^{-1}$ for the  [0.5-2] keV vs. 3.6 $\mu$m and [0.5-2] keV vs. 4.5 $\mu$m, respectively.
On the right side of the field we obtain  P$_{IR,X}$=(7.9$\pm{2.7})\times$10$^{-20}$\flux nW m$^{-2}$ sr$^{-1}$
and (5.0$\pm{1.8})\times$10$^{-20}$\flux nW m$^{-2}$ sr$^{-1}$ for [0.5-2] keV vs. 3.6 $\mu$m and [0.5-2] keV vs. 4.5 $\mu$m, respectively.
While the lower number of independent Fourier elements makes the measurements in the two smaller sub-fields less significant, the amplitudes
of the cross-power spectra are consistent when measured in different parts of the field.

\section{Discussion\label{sec:interpretation} }

The source-subtracted CIB fluctuations are made up of two components: 1) small scales ($\lsim 20^{\prime\prime}$
are dominated by the shot-noise from all sources (known and new) below the removal threshold, while 2) the larger
 angular scales reflect CIB fluctuations produced by the clustering of the new populations \citep{kamm3}. Thus the coherence
 between the two components of the fluctuations may be different depending on the different common levels of the populations
 producing the two terms. However, it is the larger scales, where the cross-power is due to clustering of the new populations common
 to both IR and X-ray emissions which are of greatest interest to interpret here.

At large angular scales ($\gsim 20''$), where the clustering term dominates the CIB fluctuations spectrum, 
the coherence between 4.5\um\ and [0.5-2] keV is ${\cal C}\sim 0.02-0.05$. 
The coherence at 3.6 \um\ versus [0.5-2] keV is consistent with these values, although the cross-power is less statistically significant.

Because the measured cross-power between the 4.5\um\ and X-ray data is highly positive,
we plot in Fig. \ref{fig:coherence} the CIB fluctuations produced by sources common to both the source-subtracted CIB at 4.5 $\mu$m and  the [0.5-2]keV CXB,
 i.e. $P_{\rm CIB, common} = P_{\rm IR,X}^2/P_{\rm X} \equiv {\cal C} \times P_{\rm IR}$.
 This assumes that {\it all} of the CXB power spectrum is produced by these sources and
 implies a {\it lower} limit on the CIB fluctuations contributed by the common sources. The figure shows
 that $\gsim15-25\%$ of the CIB power spectrum can be accounted for by these sources. The
  rightmost panel of Fig. \ref{fig:coherence} gives a similar plot for the minimal contribution,
   $P_{\rm CXB, common} = P_{\rm IR,X}^2/P_{\rm IR} \equiv {\cal C} \times P_{\rm X}$, to the CXB from the common populations.
If we consider that the CXB power  spectrum may be contaminated by the foreground contribution of the Galaxy
that at the moment, is not possible to model, the fraction of CIB power produced by X-ray sources quoted above
must be considered as a {\it lower limit}.
 We must emphasize that the ``common population'' does not necessarily imply that the corresponding parts of the CIB and 
 CXB are produced by the same physical sources emitting at both IR and X-rays. With the map resolution of a few arcsec 
 we cannot resolve the individual point sources, especially if they are at high $z$. This is further amplified since 
 the Gaussian regime of the X-ray maps is reached at angular scale of $\simeq 10^{\prime\prime}$ which subtend
  linear scale of $\sim 0.1h^{-1}$Mpc at $z\sim$1 and this defines 
  the scale of the individual ``objects'' in our analysis and the discussion below.
   Thus we cannot resolve whether the IR and X-ray emitters are one and 
   the same or whether they are different sources that share the same environment 
at the relevant angular scales.
Moreover, from the  amplitude of the cross-correlation signal  itself it is not possible
to directly determine if the signal is produced by a single population of sources, or 
if it is produced by different populations sharing the same environment.

\subsection{Galactic, solar system, and instrumental foregrounds}
We begin by considering possible non-cosmological contributions to the detected cross-correlation.
 In the IR bands, the most significant foreground source of fluctuations would come from the Galactic cirrus emissions. 
 Yet, it was demonstrated by \citet{k12} that the bulk of the 
 measured 3.6\um\ and 4.5\um\ power cannot be produced by cirrus. 
  Cirrus emission is produced by dust in cold neutral and molecular clouds which cannot emit X-rays, but can be effective absorbers of the 
  soft X-ray background leading to a {\it negative} contribution to the positive cross-power that is measured \citep{wang,snow}. 
The Galactic X-ray emission from the hot phase of the ISM could play a role 
in the cross power, but there are several factors that limit its contribution to the 
cross power: 1) It is well known that the hot ISM mostly emits soft X-rays with energy $<$1 keV \citep{frei}, and thus
 would be relatively weak even in the [0.5-2] keV band; 2) the Galactic X-ray background shows clustering on scales on the  order of one degree,
which is larger than the scales of interest here; and 3) the dust producing the IR emission is in the cold phases of the ISM, and thus should be anti-correlated
with the hot ISM, leading to negative cross-power. 

Very faint Galactic stars could, in  principle, contribute to the  cross-power. However at high
Galactic latitudes \citet{lehmer} showed that stars are a negligible component of the 
unresolved CXB. Moreover, the high level of isotropy of the CIB fluctuations
works against the hypothesis of any Galactic sources as the possible sources of CIB fluctuations. 

Other possible sources of contamination discussed by \citet{k12} are zodiacal
light and instrumental stray light whose contributions to CIB fluctuations which were demonstrated to be negligible. 
At a low level, the IR zodiacal light may correlate with Solar System X-rays generated by solar wind charge
exchange (SWCX). However, SWCX primarily produces very low surface brightness \ion{O}{7} emission around
0.54 keV where the Chandra effective area is very low and should not produce such a high signal.
Moreover SWCX emission is time dependent and therefore since {\em Spitzer} and {\em Chandra}
observed the field at different epochs, the signals are unlikely to show a correlation
due to solar system effects.

To summarize, our analysis points to an extragalactic origin
of the positive cross-power spectra between the soft X-rays and the 3.6 and 4.5 
$\micron$ background fluctuations.

\subsection{Extragalactic  populations}

Several classes of extragalactic populations could contribute to the observed
CXB-CIB cross-correlation.
Below we briefly discuss the most obvious candidates for the emissions. More detailed interpretation
will be worked out elsewhere, although it already appears that some of the candidates can be safely ruled out. 
For proper interpretation of the measured CIB-CXB correlation it is important to reiterate the limits imposed from
the IR analysis itself. The sources in the IRAC maps used here are removed down to the shot-noise $P_{\rm SN} \simeq 30$ nJy$\cdot$\nwm\ \citep{k12},
which is equivalent to sources removed to magnitudes of $m_{\rm AB}\simeq 25-25.5$ \citep{kamm1,kari}. Therefore, for this discussion
we adopt as the flux limit of $S_{\rm lim}\sim 300$nJy at the IRAC bands. Thus in order to account for the measured CIB
 fluctuation of $\delta F\sim 0.05-0.1$\nwm\ these sources must have projected angular number
 density $n\gsim \delta F/S_{\rm lim}\times (\delta F/F)^{-1}$ where $F\sim nS$ is the CIB level produced by them.
 The remaining CIB sources below the threshold would have to have $n \gsim (0.3-0.4) [(S_{\rm lim}/300{\rm nJy})(\delta F/F)]^{-1}$arcsec$^{-2}$ in
 order to explain the observed CIB at 3.6-4.5 \um.  Only the sources that can produce highly non-linear CIB fluctuations, $\delta F/F \gg 1$ all the 
 way to sub-degree scales, can have projected number density significantly lower than this. The CIB from such sources would,
  however, then exhibit a clear void-cluster CIB pattern contrary to what we see in the CIB maps.
Consequently our measurements indicate that {\it in order to explain the 
 detected cross-correlation the sources producing them would have to account for $\gsim \sqrt{{\cal C}}$ or $\gsim 15-25\%$ of the CIB signal
 and be abundant enough to reproduce the required number density   while accounting for the remaining CXB fluctuation.} 


\begin{figure}[h!]
\centering
\includegraphics[angle=90,width=.45\textwidth]{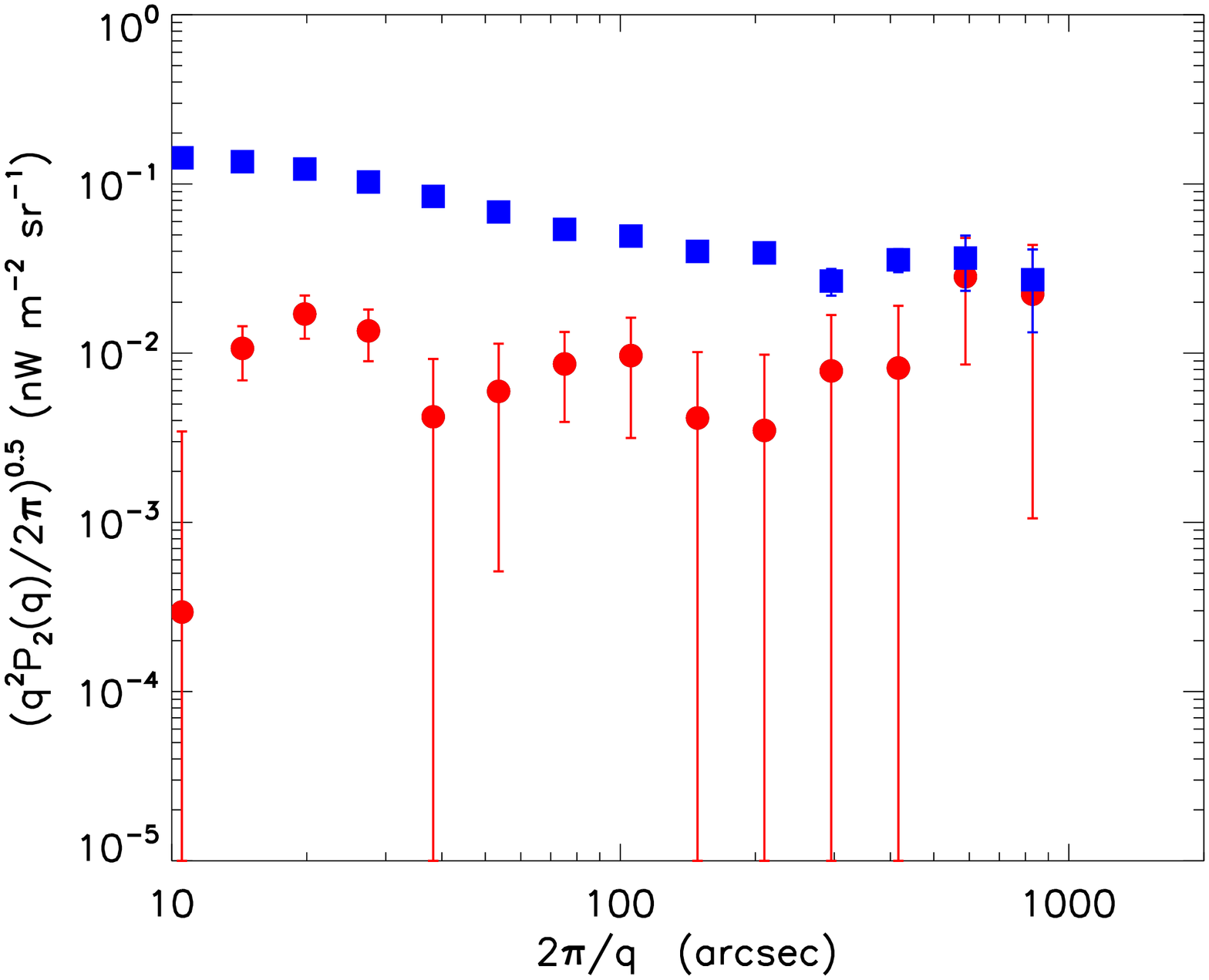}
\includegraphics[angle=90, width=.45\textwidth]{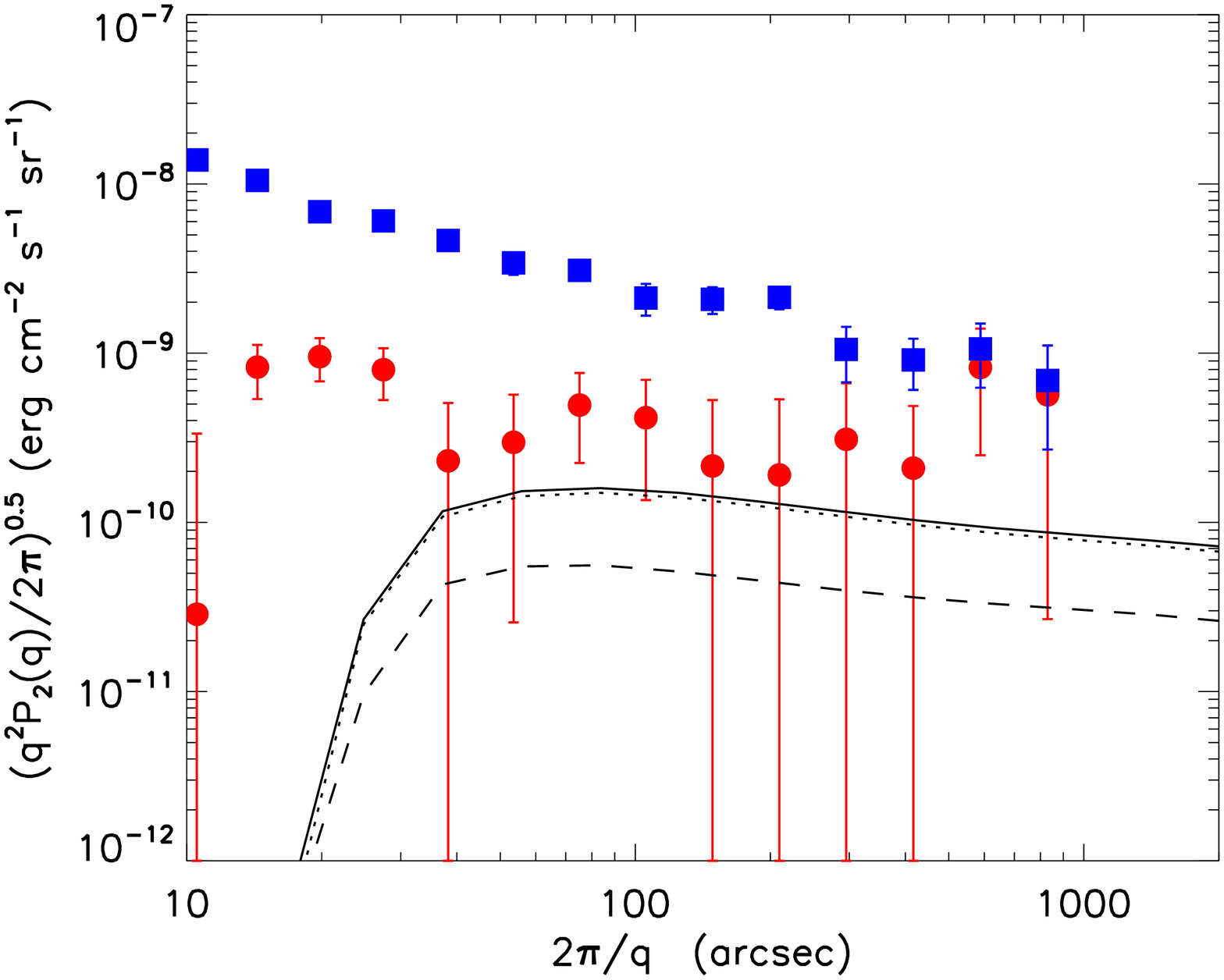}
\caption[]{ ($left$): The IRAC 4.5 \um\ fluctuation power spectrum (blue squares) compared with the 4.5 \um\ power from accreting sources (red circles).
($right$): The Chandra [0.5-2] keV fluctuation power spectrum (blue squares) compared with the power-spectrum of 
X-ray sources correlating with IR 4.5 \um\ CIB. (red circles). The dotted line is the expected upper-limit for power spectrum for 
remaining X-ray normal galaxies with $m_{AB}>25$ at 4.5 $\mu$m and $X/O<-1$.  
The dashed line is the expected upper-limit for power spectrum for 
remaining AGN with m$_{AB}>$25 at 4.5 $\mu$m and $X/O<0$.
The continuous line is the sum of the AGN and galaxies fainter than $m_{AB}\sim25-25.5$.
 } \label{fig:coherence}
\end{figure}


\subsubsection{Diffuse gas in clusters and WHIM}
As mentioned in the introduction, the sources of the unresolved soft X-ray CXB power
are mostly galaxy groups and the putative WHIM \citep[Warm Hot Intergalactic Medium;][]{cen}. However the mass limit of our source
detection allows us to exclude from our analysis extended sources with mass $M>10^{13}\Msun$.
Galaxy cluster scaling relations \citep[see e.g.,][]{pratt} ensure in this case that these sources
have a low $kT$ (i.e. $\lsim1$ keV).
Any sources correlated with clusters of galaxies at even marginally high $z$, would 
require the gas to be at temperatures shifted upward by a factor of $(1+z)$, making the origin in 
this component even less likely, doubly so since the clusters/groups are expected to have colder gas at the early times.
We therefore conclude that such a population cannot be responsible for
these measurements. This is further confirmed by the fact illustrated in Fig. \ref{fig:irps} which 
shows that the additional X-ray masking, which includes the resolved X-ray
 sources, does not have any noticeable effect on the measured CIB fluctuations.
Similarly, the WHIM, although it
has never been significantly detected in emission, is expected to show a typical emission
line dominated spectrum. Most of the emission is produced by H-, He-like
O and Ne like ions, which emit at energies below $<$ 700 eV. 
We also note that the diffuse sources producing  the CXB peak at $z\sim0.1$. If the observed
cross-correlation arose at that redshift, then the IR sources would be a population of 
still undetected numerous low luminosity \citep[i.e. with L$\sim$10$^7$ $L_\odot$,][]{kamm3} galaxies,
which further weakens this low-$z$ hypothesis.

\subsubsection{X-ray emission in remaining ``normal'' galaxies}
X-ray binaries and supernova remnants are the main sources of X-rays from normal galaxies.
 It has been shown
that this population emits X-ray with a typical spectrum $n(E) \propto E^{-2}$ \citep{rana}.
Therefore their emission could contribute to the whole energy range sampled here.
Although it is not straightforward to determine the effective X-ray flux
of galaxies with IR counterparts at $m_{\rm AB}\gsim25-25.5$, a useful tool to determine the effective X-ray brightness of
galaxies below this magnitude limit is the X-ray to optical (2500\AA)  ratio ($X/O$). In fact, it has been shown that for X-ray sources, the X-ray to
 optical/IR flux ratio assumes well defined values according to the nature of the sources.

 The $X/O$ is defined as $X/O = \log( f_X / f_{\rm opt} )=\log( f_X )+C+ m_{Vega}/2.5$.
For the 4.5\um\  vs [0.5-2] keV band the constant $C$ has a value of $\sim$7.53 \citep{civ}.
For observations with a depth comparable with ours, the value of $X/O$ for normal galaxies is $X/O \ll-1$ \citep{xue}.
Thus X-ray galaxies with IR counterpart with $m_{\rm AB}\gsim25-25.5$ should have the [0.5-2] keV flux $f<3-4\times 10^{-18}\flux$,
which is about one order of magnitude below the flux limit of the 4Ms CDFS \citep{xue}
and $\sim20$ times fainter than our limit for the EGS field.

In order to determine if these faint X-ray sources could produce the observed fluctuations
we adopted the recipe of \citet{cap12} and computed the expected CXB fluctuations angular auto-power spectra produced
 by the clustering component of these sources, i.e. the power in the X-ray bands from normal galaxies below
$z\sim 7.5$ and $f([0.5-2]{\rm keV})<3-4\times 10^{-18}\flux$. This contribution, which is of the order of 2-3\% of 
the total CXB power, is shown in Fig. \ref{fig:coherence}, right and is systematically small compared to the measured power on scales
$20''-200''$.

Additional evidence against a significant contribution  of normal galaxies is: 1) 
the shot-noise component in the CIB fluctuations on small scales, which is dominated by 
the undetected normal galaxies, appears uncorrelated with that in the CXB as is shown
 by the drop in the correlated power at the smallest scales (see  Fig. \ref{fig:coherence}), and 2) on large
 scales, which are dominated by the clustering component, the {\it minimal} 
 CIB fluctuation shown in Fig. \ref{fig:coherence} appears
  larger than the normal galaxy component reconstructed by \citet{kari} as displayed in 
  the lower right of Fig. 9 of \citet{k12}.

Thus {\it normal} galaxies could  be responsible only for a small part
of the observed signal. 

\subsubsection{Remaining known AGNs}
AGNs are characterized by  strong IR emissions due to reprocessing gas in the nuclear regions (torus)
\citep{elvis}, and/or the contribution of star forming processes in the host galaxy.
The contributions to the CIB from these sources at intermediate $z\sim 2-4$ would arise from the 
 IR bump produced by hot dust with maximum temperature of $\sim 10^3$ K.

One should therefore consider whether known AGNs can be 
responsible for the observed
cross-correlation between the source-subtracted CIB and CXB.  
A critical point in estimating their contribution to the
measured cross-power is that the signal is produced by sources below 
the IR flux of $S_{\rm lim}\simeq 300$ nJy
 at 3.6 and 4.5 \um\ which is fixed by the measured shot-noise level remaining in the CIB maps \citep{kamm1,kari,k12}. Treister et al (2004, 2006)
 conducted a detailed Spitzer/IRAC-based census and modeling of the Type I and II AGNs in the GOODS region
 and their results show that one expects the total number density of Type I and II AGNs to be $n_{\rm AGN}\simeq 6000$deg$^{-2}$ at
 the IR fluxes below $S_{\rm lim}$. The  CIB flux at 3.6 and 4.5 \um\ from the undetected  AGNs is then
 $I_{\rm AGN}\simeq n_{\rm AGN}S_{\rm lim}=6\times 10^{-6}$MJy/sr or $F_{\rm AGN} = 0.004$  nW/m$^2$/sr at 4.5 $\micron$. Thus if the AGNs were to produce the measured CIB
 signal of $\delta F \simeq 0.05$ nW/m$^2$/sr at 4.5\um\ at sub-degree scales \citep{k12}, with their X-ray emissions accounting
 for the observed cross-power, the resultant CIB would have to have highly non-linear fluctuations on scales between
 1$^\prime$ and 1$^\circ$ with $\delta F/F\gsim 10$. 
A possibility would be that the signal could be produced by a population of faint  CIB galaxies correlating with highly 
biased high-$z$ AGN. 

 A new study \citet{xue12} reported a significant
  contribution to the unresolved CXB ($\sim$25\%) at [6-8]keV from {\it highly absorbed} AGN  
  with very faint optical counterpart ($25<m<28$ at 0.85\um). The quoted result is at 3.9$\sigma$ 
  significance at [6-8] keV, while these populations are not detected below 4 keV. Thus they
  cannot be responsible for the observed effect since the correlated maps are all at energies effectively much below
   6 keV. We further emphasize that only sources with 3.6 and 4.5\um\ fluxes below $S\sim 300$ nJy contribute to the
    measured fluctuations.
Obscured AGN are the most abundant sources among faint AGN \citep{has08}. They typically
show very hard spectra and weak  X-ray emission below
$\simeq$ 3-5 keV. Since we did not detect a hard X-ray cross-power spectrum, these sources, if AGNs, would be either
Type-I sources or high-$z$ ($z>$2-4) obscured AGN, with their primary power-law component 
redshifted to the [0.5-2] keV band. 

Cappelluti et al (2012) calculated the expected clustering component of the angular auto-power spectrum produced by
AGNs with IRAC 4.5\um\ counterparts with m$_{AB}>$25-25.5. The median $X/O$ value for X-ray selected AGN is  $\sim$0 \citep{xue,civ}. 
Therefore, in order to produce the observed cross-correlation they should have [0.5-2] keV fluxes $<3-4\times 10^{-17}\flux$.
By using the recipe of \citet{cap12} we evaluated their expected angular auto-power  under the assumption
that they lie at $z<$7.5. Our prediction is shown in the right panel of Fig. \ref{fig:coherence}. 
Its amplitude is of the order 7-8\% of the total CXB fluctuations observed here. When added to the {\it normal} galaxies
component this adds up to 10-11\% of the total CXB fluctuations which is about 50\% of the observed lower limit. 


\subsubsection{New high-$z$ populations}

Although no direct measurement of the redshift of the source-subtracted CIB fluctuations is yet available,
there is now a significant body of evidence that the fluctuations may originate at early times of the Universe's evolution:
1) The measured amplitude of the fluctuations cannot be accounted for by the low-luminosity end of
the distribution of ``ordinary''/known galaxies \citep{kamm1,kari}. 2) There are no correlations between
the source-subtracted CIB maps at {\it Spitzer} wavelengths and {\it HST}/ACS data out to 0.9 \um, which points to $z> 7-8$
 for the populations producing the large scale excess signal unless the latter comes from new, and so far unobserved,
 very faint and more local populations at AB mag $\gsim 28$ which have escaped the ACS detection \citep{kamm4}.
3) The pattern of the fluctuations is inconsistent with that of the galaxy populations at
recent times, and is consistent with the $\Lambda$CDM-distributed sources at high $z$ \citep{kamm3,kamm4,k12}.
4) The colors of the fluctuations from 2 to 4.5 \um\ are consistent with very hot sources at high-$z$ \citep{akari}.

If the X-ray signal comes from sources at high redshifts we clearly do not see 
direct stellar photospheric emissions, since massive metal-poor stars have $T\simeq (9-10)\times 10^4$K \citep{schaerer1},
which is not hot enough to contribute to emissions in the observed X-ray bands extending to 7 keV.
Instead, if the signal originates from these sources, the contribution to the
CXB signal would originate from thermal emission of the gas in accretion disks.

We measure a coherence of ${\cal C}\sim0.02-0.05$. So if the BHs amongst the sources responsible for the measured source-subtracted
CIB fluctuations produce the entire X-ray signal, they should account for $\simeq 15-25\%$ of the signal produced at 4.5 \um.
If they contribute only a fraction of the X-ray fluctuations, their IR contribution would be even {\it higher}, but the measured
cross-power suggests that $\zeta_{\rm X} \gsim 15-25\%$. At the lower limit of $\zeta_{\rm X}$ the accreting sources
 (BHs?) would need to account for the {\it entire} CIB signal at 4.5 \um\ (i.e. $\zeta_{\rm IR} = 1$).

If the high-$z$ sources are responsible for the detected cross-power, these early X-ray sources were present when the Universe was
still partly neutral. 
Unlike UV photons, X-rays have the capability of
multiple ionizations.  If the sources responsible for the observed cross-correlation
are at high-$z$, we are observing correlations between the visible ($<$4500\AA) and
hard X-ray output of primordial accreting sources.
Several authors suggested that early black hole X-ray feedback
was necessary to reionize the Universe \citep{madau,ricotti1,ricotti2,giallongo}, and these results may suggest that  the early universe
was significantly irradiated by hard X-rays which  could have contributed to the reionization
of the Universe.

\subsubsection{New low-$z$ sources}

If the CIB fluctuations that we have uncovered in {\it Spitzer} data arise from new populations at lower redshifts, 
say $z\sim 2-4$, they would have to originate in low mass (faint) system 
in order to account for the lack of correlations between {\it Spitzer} maps and ACS 
sources measured in \citet{kamm4}. The cross-power spectrum of CXB and CIB then requires that
 such a model would have to explain the existence of the significant BH emitters among 
 these populations, sufficient to account for the observed contribution to the $>$[0.5-2]$(1+z)$ keV band emissions. 
An example of the intermediate $z$ sources to explain the Kashlinsky et al (2012) sub-degree CIB measurements
has been proposed in \cite{cooray2012b} as intergalactic stars stripped of their haloes at $z\sim 2-4$. Although it is not clear whether that proposal 
 can satisfy the measurement in KAMM4, we note that it is clearly problematic in light of the results discovered here.

\section{Summary}
In this paper we have presented the discovery of 
the statistically significant correlation between the 3.6\um\ and 4.5\um\ source-subtracted CIB fluctuations with the [0.5-2] keV CXB. 
Here we summarize our main results:

\begin{itemize}
\item We detected a 3.5$\sigma$ to 5$\sigma$ significance cross correlation signal
between the 3.6 \um\ and 4.5 \um\ source-subtracted CIB fluctuations and the Chandra-based [0.5-2] keV CXB
fluctuations after masking X-ray detected sources and IRAC sources down to m$_{AB}\gsim$25-25.5.
\item With this dataset  we do not find statistically-significant cross-power signal with the CXB at the harder X-ray Chandra bands ([2-4.5] keV and [4.5-7] keV).\item The cross-power appears to be of  extragalactic origin. 


\item This result presents an important step in identifying the nature of the populations producing the source-subtracted 
CIB fluctuations discovered in {\it Spitzer} data. These populations must contain a significant 
population of BHs which account for at least $\sim 15-25\%$  of the measured CIB signal.

\end{itemize}

  

\acknowledgements
NC, AK, RA acknowledge NASA Chandra Archival research grant No. AR2-13014B
for partial support. NC acknowledges the INAF fellowship program. 
We acknowledge financial contribution from the agreement ASI-INAF I/009/10/0.
NC acknowledges
 the Della Riccia foundation for  partially funding this project.

{\it Facilities:} \facility{Chandra}, \facility{Spitzer}, \facility{XMM-Newton}.

\end{document}